\documentclass[dvips]{article}

\newcommand{\cf}{\textit{cf}.\ }
\newcommand{\eg}{e.g., }
\newcommand{\etal}{\textit{et al}.\ }
\newcommand{\ie}{i.e., }

\newcommand{\vs}{\textit{vs}.\ }

\newcommand{\sect}[1]{Section \ref{s:#1}}    

\newcommand{\eqn}[1]{Eq.\ (\ref{e:#1})}      
\newcommand{\fig}[1]{Fig.\ \ref{f:#1}}       
\newcommand{\tbl}[1]{Table \ref{t:#1}}

\newcommand{\code}[1]{\texttt{#1}}

\newcommand{\bsym}[1]{\mbox{\boldmath $#1$}} 

\newcommand{\ale}{\raisebox{-.3ex}{$\stackrel{<}{\scriptstyle \sim}$}}

\def\paper#1 #2 #3 #4 #5 #6 {#1 #2. #3. \textit{#4} \textbf{#5}, #6.}
\def\inpress#1 #2 #3 #4 {#1 #2. #3. \textit{#4}, in press.}
\def\submitted#1 #2 #3 #4 {#1 #2. #3. \textit{#4}, submitted.}
\def\azpress#1 #2 #3 #4 #5 #6 {#1 #2. #3. In \textit{#4} (#5, Eds.),
  pp.\ #6. Univ.\ of Arizona Press, Tucson.}

\usepackage[final]{graphics}

\newcounter{fignum}

\newcommand{\putfig}[3]{ 
  \clearpage
  \refstepcounter{fignum}
  \label{f:#1}
  \begin{center}
    Figure \thefignum
    \vfill
    \resizebox{#2}{!}{\includegraphics{#3.ps}}
    \vfill
  \end{center}
}

\newcommand{\figcap}[1]{\item{\bfseries Figure \ref{f:#1}:}}

\renewcommand{\baselinestretch}{1.5} 
\newcommand{\hugeskip}{\bigskip\bigskip\bigskip}

\begin{document}
%
%
\thispagestyle{plain}
\setlength{\footskip}{0pt} 
\begin{center}
  {\large\bf Direct $N$-body Simulations of Rubble Pile Collisions} \\
  \hugeskip
  \textbf{Zo\"e M. Leinhardt}$^\dag$ \\
  \textbf{Derek C. Richardson}$^\dag$ \\
  \textbf{Thomas Quinn}$^\dag$ \\
  \bigskip
  $^\dag$Department of Astronomy \\
  University of Washington \\
  Box 351580 \\
  Seattle, WA 98195-1580 \\
  Tel: (206) 543-8989 \\
  Fax: (206) 685-0403 \\
  E-mail: \texttt{zoe@astro.washington.edu} \\
  \bigskip
  Printed \today \\
  \bigskip
  Submitted to \textit{Icarus} \\
  \vfill
  30 manuscript pages including 4 tables \\
  8 figures including 1 in color \\
  \hugeskip
\end{center}
\begin{description}
\item{\itshape Key Words:} Asteroids, collisional physics, impact
  processes, planetary formation, planetesimals.
\end{description}
\newpage
\begin{flushleft}
  Running page head: Simulations of Rubble Pile Collisions \\
  \bigskip
  Please address all correspondence to: \\
  \bigskip
  Zo\"e M. Leinhardt \\
  Department of Astronomy \\
  University of Washington \\
  Box 351580 \\
  Seattle, WA \\
  98195-1580
\end{flushleft}
\newpage
\section*{Abstract}

There is increasing evidence that many km-sized bodies in the Solar
System are piles of rubble bound together by gravity.  We present
results from a project to map the parameter space of collisions
between km-sized spherical rubble piles.  The results will assist in
parameterization of collision outcomes for Solar System formation
models and give insight into fragmentation scaling laws.  We use a
direct numerical method to evolve the positions and velocities of the
rubble pile particles under the constraints of gravity and physical
collisions.  We test the dependence of the collision outcomes on
impact parameter and speed, impactor spin, mass ratio, and coefficient
of restitution.  Speeds are kept low ($< 10$ m s$^{-1}$, appropriate
for dynamically cool systems such as the primordial disk during early
planet formation) so that the maximum strain on the component material
does not exceed the crushing strength.  We compare our results with
analytic estimates and hydrocode simulations.  We find that net
accretion dominates the outcome in slow head-on collisions while net
erosion dominates for fast off-axis collisions.  The dependence on
impact parameter is almost equally as important as the dependence on
impact speed.  Off-axis collisions can result in fast-spinning
elongated remnants or contact binaries while fast collisions result in
smaller fragments overall.  Clumping of debris escaping from the
remnant can occur, leading to the formation of smaller rubble piles.
In the cases we tested, less than 2\% of the system mass ends up
orbiting the remnant.  Initial spin can reduce or enhance collision
outcomes, depending on the relative orientation of the spin and
orbital angular momenta.  We derive a relationship between impact
speed and angle for critical dispersal of mass in the system.  We find
that our rubble piles are relatively easy to disperse, even at low
impact speed, suggesting that greater dissipation is required if
rubble piles are the true progenitors of protoplanets.

\newpage
%
%
\section{INTRODUCTION} \label{s:intro}

There is growing interest in understanding the dynamics of collisions
between small bodies in the Solar System.  Typically such collisions
are divided into two regimes: those dominated by material strength and
those dominated by self-gravity (Holsapple 1994).  The transition from
the strength to the gravity regime may occur at body sizes as small as
250 m for silicates (Love and Ahrens 1996).  In this paper we present
numerical results from simulations of collisions in the gravity
regime.  Our experiments are primarily concerned with low-speed
collisions between equal-mass, km-sized ``rubble piles'',
gravitationally bound aggregates of loose material.  We believe that
these experiments will shed light on the collisional dynamics of the
protoplanetary disk when typical encounter speeds are comparable to
the surface escape speed (about 1 m s$^{-1}$ for km-sized
planetesimals of 2 g cm$^{-3}$ bulk density).

\subsection{Definitions}

We begin with definitions of terms frequently encountered in the
context of binary collision experiments.  Typically in the literature
one impactor (the larger one) is stationary and is considered to be
the \emph{target} while the other (the smaller one) is moving and is
called the \emph{projectile}.  In our experiments, the impactors are
comparable in size and are both in motion, so we generally do not
distinguish between a target and a projectile.  Most laboratory
experiments involve solid targets that possess tensile strength, so
the outcome is measured in terms of the extent of \emph{disruption} or
\emph{shattering} of the target.  A \emph{critical} or
\emph{catastrophic} shattering event is one in which the largest
post-impact fragment (the \emph{remnant}) has 50\% of the target mass.
Following a recently adopted convention in the literature (Durda \etal
1998), we use $Q^\star_S$ to denote the kinetic energy per unit target
mass to achieve critical shattering.  A rubble pile, by definition,
has no tensile strength, so $Q^\star_S$ is effectively zero.  However
a rubble pile can still be disrupted or shattered in the sense that
one or more of the component particles becomes separated from the rest
for at least an instant.

For collisions in free space, fragments or particles are said to be
\emph{dispersed} if they attain positive orbital energy with respect
to the remnant.  Hence a critical or catastrophic dispersal is one in
which the largest remnant is left with 50\% of the original target
mass after the remaining material has dispersed to infinity.  The
energy per unit target mass to achieve this is denoted by $Q^\star_D$.
In our experiments, since we do not distinguish between a target and a
projectile, $Q^\star_D$ refers to the energy per unit \emph{total}
mass, in the center-of-mass frame, needed to critically disperse the
entire system.

Finally, we define \emph{erosion} to mean permanent removal of mass
from a body, and \emph{accretion} to mean permanent retention of
mass.  In the context of our experiments, net erosion means that one
body (the largest if the impactors are of unequal mass) had less mass
at the end of the run than it started with.  Net accretion means it
had more mass at the end.

\subsection{Motivation} \label{s:mot}

Many asteroid characteristics are inconsistent with monolithic
configurations.  Recent observations by the Near Earth Asteroid
Rendezvous spacecraft of Mathilde, a 53-km C-class asteroid, are
particularly suggestive.  First, Mathilde's largest crater is
enormous: it has a diameter of 33.4 km, almost 7 km larger than the
asteroid's mean radius (Veverka \etal 1997).  Numerical hydrocode
simulations and laboratory experiments strongly suggest that in order
for Mathilde to have survived the impact that formed such a
substantial crater, the asteroid must be made of some material that
does not efficiently transmit energy throughout the body (Housen and
Holsapple 1999).  Second, Mathilde has a remarkably low density of 1.3
g cm$^{-3}$ (Yeomans \etal 1997), about one third the average value
for the chondritic meteorites that are thought to originate from
C-class asteroids (Wasson 1985).  Such a low density suggests that
Mathilde is highly porous.  If true, the voids in the material could
impede the transmission of energy from a collisional shock wave and
allow a rather weak body to survive an otherwise catastrophic impact
event (Asphaug \etal 1998, Housen and Holsapple 1999).

In addition to Mathilde, the surfaces of 243 Ida, 951 Gaspra, and
Phobos show several sizable craters that have diameters on the order
of the mean radius of the body (for references, see Richardson \etal
1998, hereafter Paper I).  As in the case of Mathilde, the energy
necessary to create craters of this size would disperse or disrupt the
original body if it were solid (Asphaug \etal 1998).

Further evidence for the prevalence of rubble piles comes from
asteroid spins.  In a sample of 107 asteroids smaller than 10 km in
diameter, Harris (1996) found that the spin period distribution
truncates at fast spin rates, where rubble piles would start to fly
apart.\footnote{At least one asteroid spinning faster than this limit
  has since been discovered (Ostro \etal 1999), but its small size
  ($\sim$ 30 m) puts it comfortably in the strength regime.}

One explanation for the observed characteristics of these asteroids
and their craters is that they are rubble piles.  Although rubble-pile
configurations are more susceptible to disruption by tidal forces than
monolithic configurations (Paper I), there is increasing evidence that
rubble piles have a higher impact strength (Ryan \etal 1991, Love and
Ahrens 1996, Asphaug \etal 1998).  There are two scenarios for
creating a rubble-pile asteroid: 1) the asteroid is initially one
solid body of material and is rubblized over time by multiple impacts;
2) the rubble-pile configuration of the asteroid is primordial.
Regardless of how rubble-pile asteroids are formed it is interesting
to investigate how they interact and evolve in the Solar System.

\subsection{Laboratory Experiments: Strength \vs Gravity} \label{s:labexp}

Ryan \etal (1991) presented results from a laboratory study of impacts
into weak inhomogeneous targets.  Due to practical limitations they
used $\sim$ 0.5-cm targets of gravel and glue.  As a result, their
specific experimental results are firmly rooted in the strength
regime.  However, the most general conclusion that the group arrived
at from dropping, crashing, and shooting at the gravel aggregates was
that the relatively weak targets have a surprisingly high impact
strength.  In other words, it took a large amount of energy (at least
$Q^\star_S = 2.6 \times 10^5$ J m$^{-3}$) to critically disrupt or
shatter the target such that the largest remnant was one half the mass
of the original object.  The nonuniformity of the target causes a
greater fraction of the impact energy to dissipate thermally;
therefore, the collisional shock wave is more efficiently absorbed by
the target.

Laboratory experiments on Earth to investigate directly the
collisional dynamics of the gravity regime are difficult to conduct
since the target size necessary to reach this regime is impractically
large.  Instead, overpressure and centrifuge techniques have been used
to artificially simulate the gravity regime in the laboratory.  In an
overpressure experiment, Housen \etal (1991) used nitrogen gas at
various pressures to mimic the lithostatic stress felt inside a large
target.  At these pressures they were unable to carry out true impact
tests so they used a buried charge instead of a projectile.  As the
pressure was increased, the size of the largest remnant after each
explosion also increased, indicating a transition from the
strength-dominated regime to the pressure-dominated regime.  Housen
\etal (1991) argued that the pressure regime was analogous to the
gravity regime and extrapolated a scaling law for the gravity regime
from the overpressure data.  This laboratory study has two important
drawbacks: 1) by using a buried charge the experiment does not model
the actual surface dynamics of an asteroid during an impact; 2) the
gas overpressure is not an r$^{-2}$ force law.  They were able to
reach a regime in the laboratory that was not dominated by the
strength of the material, but it is unclear whether the gravity-regime
scaling law derived from the overpressure data is valid.

In a centrifuge experiment, Housen and Holsapple (1999) were able to
conduct true impact tests by firing a small projectile (a polyethylene
cylinder 0.65 cm in diameter) from a gas gun strapped to the arm of a
centrifuge.  They positioned a porous target (composed of quartz sand,
perlite, fly ash, and water) at the end of the arm.  The centrifuge
was used to mimic the gravitational force at the surface of a much
larger body.  The use of the centrifuge introduces second-order
complexities due to the Coriolis force and the field orientation in
general at the surface of the cylindrical target (though this is only
really a problem in the event of high ejecta trajectories).  In
addition, the flat surface of the target may subtly affect crater
morphology.  Nonetheless, this experiment showed that porous targets
in the gravity regime are efficient at absorbing impact energy at the
surface by compacting the underlying material.

\subsection{Numerical Simulations of Collisions and Tidal Disruptions} \label{s:numexp}

Extrapolations of laboratory experiments have resulted in rough
strength and gravity scaling laws.  In order to truly understand the
collisional dynamics and evolution of large bodies, numerical
simulations are a necessity.  Love and Ahrens (1996) used a
three-dimensional ``smoothed particle hydrodynamics'' (SPH) code to
simulate high-speed catastrophic collisions.  They used various impact
speeds (3--7 km s$^{-1}$), impact angles (5--75$^\circ$), target
diameters (10--1000 km), and projectile diameters (0.8--460 km) in
order to explore a large region of parameter space.  The big targets
placed the experiments securely in the gravity regime, allowing the
researchers to treat gravity carefully and neglect the strength and
fracturing of the target completely.  Their extrapolated scaling law
for the gravity regime placed the transition from the strength to the
gravity regime at a target diameter of $250 \pm 150$ meters, much
smaller than that predicted by laboratory experiments (Holsapple
1994).  Love and Ahrens (1996) argue that since smaller asteroids are
more common than larger ones, a given asteroid is more likely to
suffer a shattering impact before a dispersing impact.  Thus, it seems
plausible that many asteroids in our Solar System are at least partial
rubble piles.

More recent simulations have had similar results.  Asphaug \etal
(1998) conducted three high-speed (5 km s$^{-1}$) impact experiments
using a solid target, a partially rubblized contact binary, and a
totally rubblized ellipsoidal target.  In each case the researchers
used a small projectile six orders of magnitude less massive than the
target.  There are three major conclusions from this study: 1) it is
much easier to disrupt a solid target than it is to disperse it---this
conclusion is evidence that it is possible to change a solid body into
a rubble pile with impacts; 2) rubble regions can insulate and block
energy from traveling through a body---in a contact binary, for
example, one end could be critically disrupted while the other remains
undamaged; 3) the fully rubblized targets efficiently localize the
energy transmitted during a collision which in turn minimizes the
damage outside the collision region and allows weak bodies to survive
high-energy impacts with much less damage than solid targets.  This
again implies that many small bodies in the Solar System may be rubble
piles.

Watanabe and Miyama (1992) used 3-D SPH code to investigate the
effects of tidal distortion and shock compression from collisional
impacts in the process of planetary accumulation.  They used two
equal-sized spherical bodies and assumed a perfect Newtonian fluid.
It is important to note that their code did not model an
incompressible fluid (their adopted polytropic indices were always
greater than zero).  As a result of experimenting with impact angle,
speed, and density gradients, they found that tidal forces can enlarge
the coalescence rate of planetesimals by almost a factor of 2.  In
addition, when the initial speed of the impactor is significantly
lower than the escape speed of the system, less than a few percent of
the total mass is lost from the system in the collision.  They did not
attempt any simulations with initial speeds in excess of 50\% of the
escape speed.

In Paper I, Richardson \etal numerically simulated the effects of the
Earth's tidal force on rubble-pile asteroids.  Unlike Watanabe and
Miyama (1992), they simulated the Earth-crossing asteroids as
incompressible fluids using a hard-sphere model.  They varied the
asteroids' speed, spin, shape, and close-approach distance.
Generally, slow-moving, close-approaching, prograde-rotating,
elongated asteroids were the most susceptible to tidal disruption.
They found several distinct classes of outcome: in the most violent
disruption cases, the asteroid was stretched into a line and
recollapsed into a ``string of pearls'' reminiscent of Comet
D/Shoemaker-Levy 9 at Jupiter; for moderate disruptions, large pieces
of the asteroid were stripped off in many cases, forming satellites or
contact binaries; the mildest disruptions resulted in little mass loss
but significant shape changes.  These various outcomes could lead to
the formation of crater chains (Bottke \etal 1997), asteroid
satellites and doublet craters (Bottke and Melosh 1996a,b), and
unusually shaped asteroids (Bottke \etal 1999).

Durda (1996) carried out simulations to study how readily satellites
form as a result of mutual gravitational attraction after the
catastrophic disruption of the progenitor.  Durda (1996) came to three
major conclusions: 1) satellites do form immediately after a
catastrophic collision; 2) contact binaries form more easily than true
binary systems; 3) the binary systems form in a wide range of size
ratios.  It is important to realize that Durda (1996) assumed a
power-law mass distribution for the catastrophically fragmented
asteroid.  The slope index used (1.833) was taken from extrapolations
of laboratory experiments.

\subsection{Implications for Planet Formation} \label{s:sims}

Traditionally, numerical simulations of planet formation use
extrapolations of impact experiments in the strength regime to model
the effects of fragmentation in planetesimal collisions (\eg Greenberg
\etal 1978; Beaug\'e and Aarseth 1990; Wetherill and Stewart 1993).
From what we have already seen, such extrapolations may give
misleading results since generally much more energy is needed to
disperse than to disrupt a planetesimal in the gravity regime.
Moreover, effects of impact angle, spin, and impactor mass ratio are
often not taken into account.  In the case of rubble piles, no
empirical model actually exists.  For example, we might expect
reaccumulation like that seen in the tidal disruption models to also
occur after the catastrophic impact of two rubble-pile planetesimals.
In this paper we aim to explore these issues by simulating collisions
between rubble-pile bodies over a wide range of parameter space and
determining the implications of the results for planet formation.  In
\sect{method} we describe our numerical method and analysis technique.
Our results are presented in \sect{results}, followed by a general
discussion in \sect{discuss}.  We give our conclusions in
\sect{concl}.

\section{METHOD} \label{s:method}

The simulation and analysis of the collisions presented here combine
numerical methods introduced in Paper I and Richardson \etal (1999),
hereafter Paper II.  The rubble pile model is an extension of the
model used for studying the tidal disruption of asteroids (Paper I).
The integration engine is an extension of the parallel tree code used
for planetesimal evolution simulations (Paper II).

\subsection{Rubble Pile Model} \label{s:rubble}

Each rubble pile in our simulations consists, at least initially, of a
fixed number of equal-size hard spheres arranged in ``hexagonal
close-packed'' (HCP) form.  The rubble piles are typically generated
by specifying the bulk semi-axes, bulk density, and approximate number
of particles (alternatively, the particle radius and/or density can be
used as independent parameters).  The generator attempts to match the
requested properties on the basis of the estimated HCP efficiency of a
sphere as a function of bulk radius or number of particles (derived
from power-law fits to our own numerical experiments).  Once the
rubble pile is constructed, the constituent particles are reduced in
size by a fixed factor (usually 1\%) and given a small random velocity
kick (no more than 10\% of the particle surface escape speed in
magnitude).  This is to facilitate attaining the initial equilibrium
(\cf \sect{ics}).  Finally, the rubble pile is tagged with a unique
``color'' so that mixing can be studied visually and statistically.

The collisional properties of the constituent particles are specified
prior to each simulation.  These include the normal and tangential
coefficients of restitution, $\epsilon_n$ and $\epsilon_t$ (\cf
Richardson 1994).  Except for certain explicit test models, these
values generally were fixed at $\epsilon_n = 0.8$ (mostly elastic
collisions with some dissipation) and $\epsilon_t = 1.0$ (no surface
friction).  Bouncing was the only possible collision outcome: no
mergers or fragmentations of particles were allowed.  The value of
$\epsilon_n$ was chosen to be consistent with Paper I and is similar
to experimentally determined values used in the literature (\eg
Beaug\'e and Aarseth 1990).  Note that in the perfectly elastic case,
particles cannot re-collapse into condensed rubble piles after a
disruption event but instead completely disperse or at best end up in
centrally concentrated swarms.  In the case of tidal disruption (Paper
I) the outcome is relatively insensitive to the choice of
$\epsilon_n$, so long as $\epsilon_n < 1$.  For the present study,
however, varying $\epsilon_n$ has a stronger effect, an issue we
explore in \sect{cofr}.  We did not include surface friction in the
present study, in order to keep the number of test cases manageable.

There are two circumstances under which $\epsilon_n$ is allowed to
change.  First, if the relative speed of two colliding
\emph{particles} is less than 10\% of their mutual escape speed (\ie
typically $\sim$ 1 cm s$^{-1}$), $\epsilon_n$ is set to unity (no
dissipation).  This is to prevent computationally expensive ``sliding
motions'' (Petit and H\'enon 1987).  Second, if the collision speed
exceeds 10 m s$^{-1}$, $\epsilon_n$ is set to 0.2 (highly
dissipative).  This is to crudely model damping through internal
fracture as the impact stress $\rho v c$ ($\rho$ = internal density,
$c$ = sound speed $\sim 10^3$ m s$^{-1}$) exceeds the ``rock''
strength ($\sim 10^7$ N m$^{-2}$).  This is not intended to be a
physically rigorous model but rather a simple mechanism to prevent
unrealistically high collision speeds.  Initial encounter speeds
between rubble piles were generally kept closer to 1 m s$^{-1}$ in any
case.  Also, particle sizes were kept roughly comparable across rubble
piles in order to minimize any strength-\textit{versus}-size biases.

It is important to note that neither rolling nor true sliding motions
are modeled in our code.  Moreover, particles cannot remain mutually
at rest in contact (\ie there are no surface normal forces).  Instead,
the constituent particles of an otherwise quiescent rubble pile are in
a constant state of low-energy collisional vibration (dictated by the
minimum sliding condition described above).  Nevertheless, such small
bounces can mimic transverse motions in an approximate sense in the
presence of shear flow, giving realistic bulk properties to the
material.  To test this we have simulated the formation of sand piles
using our collision code (with surface friction) that give reasonable
values for the angle of repose when compared with laboratory
experiments.

\subsection{Numerical Code}

Our simulations were performed using a modified version of a
cosmological $N$-body code, \code{pkdgrav} (Stadel and Quinn, in
preparation; data structures described in Anderson 1993, 1996).  This
is a scalable, parallel tree code designed for ease of portability and
extensibility.  For the parameter space study, the parallel capability
was not exploited owing to the modest number of particles in each run
(a few thousand).  However, even in serial mode, \code{pkdgrav} is
arguably more efficient than any other existing code with similar
capability.  In particular, it is superior to \code{box\_tree}, the
code used in Paper I, which could only handle a few hundred particles
in practical fashion.

A low-order leapfrog scheme is used as the \code{pkdgrav} integrator.
The comparative simplicity of this scheme is a big advantage for
collision prediction since particle position updates are linear in the
velocity term.  This means that every possible collision within the
timestep can be determined in advance and in the correct sequence.
Timesteps are smaller than in higher-order schemes for the same
accuracy, but the cost of each gravity calculation is far outweighed
by the collision search once the rubble piles are in contact, and is
comparable otherwise.  Moreover, away from collision, particle
trajectories are integrated symplectically, eliminating spurious
numerical dissipation.  For further detail and references, refer to
Paper II.

Although the collision search is relatively expensive, the scaling is
modest: ${\cal O}(N \log N)$ with particle number and linear with the
number of collisions per interval.  A typical encounter between
thousand-particle rubble piles can generate $\sim 10^8$ collisions
over the course of a run!  A balanced $k$-d tree (Bentley and Friedman
1979) is used to search for possible collisions at the beginning of
each timestep, giving the ${\cal O}(N \log N)$ dependence.  Once a
collision is performed, only particles that might be affected by the
event in the remaining interval (numbering typically $\ll N$) are
reconsidered via the neighbor search, giving the near linear
dependence on the number of collisions.  This latter enhancement is an
improvement to the Paper II code, which did not require as much
sophistication given the low collision frequency per step.  Note that
the collision search can also be performed in parallel, which proved
necessary for the large-$N$ models presented in \sect{hires} below.

\subsection{Hardware}

The parameter space models were run on a local cluster of 16 300-MHz
Intel Pentium IIs using the High Throughput Computing (HTC)
environment \code{condor} (\cf \code{http://www.cs.wisc.edu/condor/}),
version 6.0, under RedHat Linux 5.0 with a 2.0.35 Linux kernel.  The
\code{condor} system supports automatic scheduling, submission, and
restarting of jobs on shared resources, greatly simplifying
management.  A typical run required between 12 and 72 wallclock hours
to complete and each generated $\sim$ 25--50 MB of data.  Models
requiring parallel resources were run either on a local cluster of 4
433-MHz DEC Alpha PCs connected with a fast ethernet switch, or on a
local SGI Origin 200 with 4 180-MHz processors running IRIX 6.4.  Both
platforms typically achieved sustained performances of several hundred
megaflops.

\subsection{Initial Conditions} \label{s:ics}

Generation of initial conditions and analysis of results were
performed using code auxiliary to \code{pkdgrav}.  The rubble pile
generator has already been described (\sect{rubble}).  Each new rubble
pile was first run in isolation (with or without spin) using
\code{pkdgrav} until the velocity dispersion of the constituent
particles achieved a stable equilibrium.  Next a new ``world'' was
created by using a small program to position and orient any number of
equilibrated rubble piles (always two in the present study) prior to
simulation.  Spherical bodies were usually given a random orientation
in order to reduce the effect of HCP planes of symmetry.  Bulk
velocities were then applied to each rubble pile.  Other rubble pile
properties that could be changed at this point included the total
mass, bulk radius, bulk density, and color.  For the exploration of
parameter space, usually only the positions (in the form of $y$
offsets), velocities, spins, and colors were modified.  Once all the
rubble piles were in place, the world would be adjusted so that the
center of mass coincided with the origin and the velocity of the
center of mass was zero.  The output world was then read in by
\code{pkdgrav} and the simulation would begin.

To facilitate the exploration of parameter space, a series of Unix
scripts were written to generate and monitor each run.  Starting with
a given pair of rubble piles and a list of desired initial impact
parameters, speeds, and spins, the world generator was run
automatically to create the necessary initial conditions and support
files in separate run directories.  The scheduler \texttt{condor} was
then invoked to farm the jobs out to all available machines.  Analysis
was performed on the fly using a machine outside the \texttt{condor}
pool for maximum efficiency.

The choice of initial conditions was governed largely by prior test
simulations.  For the parameter space exploration, 10 values of impact
parameter $b$ and 10 values of initial relative speed $v$ were chosen
for each set of runs, where a set consisted of a fixed choice of spin
and/or offset direction (see \sect{results} for a complete description
of each model).  From the test simulations it was clear that only
about half of the possible 100 runs for each model were needed to find
the representative cases and the $Q^\star_D$ boundary.  In a plot of
$b$ \vs $v$, the important region is the lower-left triangle (see
\fig{modelA} for an example).  The $b$ and $v$ values were therefore
chosen to sample this region as finely as possible in a practical
amount of time.  Models with spin were chosen to sample representative
combinations of spin and orbital angular momentum at a fixed rotation
period.

\subsection{Coordinate System and Units} \label{s:units}

We use an inertial Cartesian coordinate system in free space for our
simulations, with the origin at the center of mass.  In the parameter
space studies, the initial motion of the colliding bodies is in the
$\pm x$ direction.  Any initial impact parameter is measured in the
$\pm y$ direction.  Most debris actually travels in directions
perpendicular to the original axis of motion (\cf \sect{iso}).

A natural unit for the impact parameter $b$ is the sum of the radii
$R_1 + R_2$ of the two (spherical) impactors.  Hence $b = 0$ implies a
head-on collision while $b = 1$ is a grazing encounter.  Note however
the true trajectories will generally be hyperbolae; no allowance is
made for this in the definition.  Since tidal effects may play an
unpredictable role anyway, we adopt the simpler definition.  In the
absence of trajectory deflection, the impact angle is then $\phi =
\sin^{-1} b$, for $b \le 1$.

The unit for the initial relative speed $v$ is more complicated.  We
chose a system in which $v = 0$ indicates no relative motion and $v =
1$ is the estimated critical speed for dispersal.  The critical speed
is found by equating the initial total kinetic energy with the
gravitational binding energy of a rubble pile made up of a spherical
and homogeneous mixture of both colliders:
\begin{equation}
  v_\mathrm{crit} = M \sqrt{\frac{6G}{5\mu R}},
\end{equation}
where $M$ is the combined mass, $G$ is the gravitational constant,
$\mu$ is the reduced mass $M_1M_2/M$, and $R$ is the radius of the
sphere that contains the combined mass, assuming the same bulk
density:
\begin{equation}
  R = \left( R_1^3 + R_2^3 \right)^{1/3}.
\end{equation}
Note that the actual speed at impact will slightly exceed $v$ due to
gravitational acceleration.

In the parameter space models, the initial separation in $x$ for all
cases was $\sim 6R$, effectively 2.5 Roche radii for the combined
mass, \ie far enough apart that initial tidal effects were negligible.
The total energy of the system was positive in all cases.  For
completeness, the speed at infinity is given by
\begin{equation}
  v_\infty = \left( v^2 v_\mathrm{crit}^2 - \frac{GM\cos\phi}{3R}
  \right)^{1/2}, \label{e:vinfty}
\end{equation}
and the speed at impact is
\begin{equation}
  v_\mathrm{impact} = \left( v_\infty^2 + \frac{2GM}{R_1 + R_2}
  \right)^{1/2}.
\end{equation}

\subsection{Run Parameters}

Most \code{pkdgrav} run parameters assumed default values for these
simulations (\cf Paper II).  However, in addition to the collision
parameters described in \sect{rubble}, the run time, timestep, and
output frequency were specified explicitly for each model.

The run time was initially 10 times the characteristic time
\begin{equation}
  t_c \sim \sqrt{\frac{x^3}{GM}} ,
\end{equation}
where $x$ is the initial separation.  In most cases this is sufficient
time for the post-collision system to reach a steady state.  Some
cases were run longer (typically a factor of 2) if necessary, on the
basis of visual inspection of animations.

The timestep for each run was set to a small value $t_0$ times a
heuristic scale factor of $1/(2v + 1)$, arrived at by trial and error
from our test runs (recall $v$ is the \emph{initial} speed, so $t_0$
is a simple constant).  The scaling ensures finer intervals for
neighbor searches in higher-speed impacts (this is necessary to avoid
missing any potential collisions).  For our runs, $t_0 = 10^{-5}$
yr/2$\pi$, or roughly 50 s.  Note that for objects with bulk density a
few g cm$^{-3}$ the dynamical time $1/\sqrt{G\rho} \sim$ 1 h,
comfortably large compared to the maximum adopted timestep.  Generally
our simulations are limited by the time needed to deal with particle
collisions, so the gravity calculations can be of higher accuracy with
little additional cost.

Finally, the output frequency was chosen so that there would be about
200 outputs per run, suitable for smooth animations and analysis.

\subsection{Analysis Method}

Much of our analysis method is similar to that presented in Paper I;
the reader is referred to that work for additional details.  The basic
strategy is to identify the largest post-collision remnant, compute
its various properties, and generate statistics for the relative
distribution of the smaller fragments.  We use a slightly different
clump-finding algorithm (\sect{clumps}), and now employ a shape
drawing technique (\sect{shapes}).  We have made other refinements
that should improve the accuracy of the analysis.

\subsubsection{Clump finding} \label{s:clumps}

The clump-finding algorithm iteratively refines guesses as to what
constitutes a rubble pile by merging groups of particles together in
bottom-up fashion.  The first guess is that every particle in the
system is its own rubble pile.  On each pass basic properties are
computed for each clump: mass, position, axis lengths, and
orientation.  Clumps are then compared in pairwise fashion.  In order
for two clumps to be merged (\ie to be considered one clump), spheres
of diameter equal to the major axes times a fixed linking scale and
centered on each clump must overlap.  If the scaled minor axes also
overlap, then the clumps are merged.  Failing that, if either body has
its center of mass in the other's scaled ellipsoid, the bodies are
merged.  Otherwise no merge occurs.  This process iterates until there
are no more mergers during an iteration.

This method is purely geometrical: gravitational groupings are not
considered.  This was done mostly because there is no natural
gravitational length scale in the present context, unlike in Paper I
where the Hill radius could be used.  However, osculating elements of
groups measured with respect to the largest fragment are still
calculated and give a good indication of the future evolution of the
system.  The present method also differs from Paper I in that it
treats each clump as an ellipsoid rather than a sphere, allowing more
refined boundaries to be drawn.  Through trial and error (visual
inspection) a linking scale of 1.1 was adopted.

\subsubsection{Shape drawing} \label{s:shapes}

During the course of the present investigation we came across some
unusual, often asymmetrical shapes following collision events.  In
order to characterize these forms, a shape-drawing algorithm was
devised.  The algorithm attempts to trace the outer surface of a given
rubble pile (either in cross section or by projection to the $x$-$y$
plane).  The resulting shape is equivalent to what would be measured
by laser beams aimed at the surface in the direction of the center of
mass.  Note that this means any outcroppings can conceal underlying
structure.  Generally such complex surfaces are not seen in our models
however (as confirmed by 3-D VRML viewing).  The projection method is
used in the parameter space plots of \sect{results}.

\subsubsection{Mixing} \label{s:mixing}

The unique color assigned to each rubble pile makes it easy to assess
visually the degree of mixing following a collision.  In order to make
a more quantitative assessment, we have constructed the following
statistic:
\begin{equation}
  f_\mathrm{mix} = 1 - \frac{1}{\sqrt{N_v}} \sum_c \left[ \sum_v
    \left( \frac{m_{c,v}}{\sum_{c'} m_{c',v}} -
      \frac{m_{c,\mathrm{world}}}{\sum_{c'} m_{c',\mathrm{world}}}
    \right)^2 \right]^{1/2},
  \label{e:mixing}
\end{equation}
where subscript $c$ denotes a color, subscript $v$ denotes a subvolume
of the rubble pile, $N_v$ is the number of subvolumes, and ``world''
refers to the entire population of particles in the system.  Note that
particle number is conserved so that $\sum_c m_{c,\mathrm{world}} =
M$, the total mass of the system.  This formula is generalized for any
number of components (colors); in the present study only two
populations are considered.  A mixing fraction of unity implies the
rubble pile contains a perfectly homogeneous mixture of the world
colors.  A value of zero means no mixing has taken place at all.

Spherical subvolumes are used to sample different regions of the
rubble pile (which itself need not be spherical).  The size of the
sample region is set so that it contains $\sqrt{N}$ particles on
average.  The center of a subvolume is chosen randomly within a
rectangular prism enclosing the rubble pile.  A new subvolume is
chosen if the region is found to contain fewer than $N^{1/4}$
particles.  Otherwise, the argument of the $\sum_v$ in \eqn{mixing} is
computed and added to the running sum.  This is repeated until
$\sqrt{N}$ subvolumes are successfully sampled.

\section{RESULTS} \label{s:results}

We now present the results of our simulations.  First we describe the
parameter space exploration which consisted of numerous runs of modest
size.  Highlights are shown in \fig{strips} where we have endeavored
to illustrate the various classes of outcomes.  Second we show the
dependence on the coefficient of restitution $\epsilon_n$ for a
particular high-energy run.  Finally we present the results of two
high-resolution cases and compare with the corresponding
moderate-resolution runs.

\subsection{Parameter Space} \label{s:params}

We divided our exploration of parameter space into three models: a
generic case as a baseline, a case with spinning impactors, and a case
with unequal-mass impactors.  Graphical summaries of these models are
given in Figs.\ \ref{f:modelA} and \ref{f:quad}, which are discussed
in detail below.

\subsubsection{Model A: Equal size, no spin} \label{s:modelA}

Model A, our generic case, consisted of two equal-size rubble piles of
1 km radius and 2 g cm$^{-3}$ bulk density.  The rubble piles were
generated and equilibrated using the process described in \sect{ics}.
Each rubble pile contained 955 identical spherical particles of 83 m
radius, so the packing efficiency was 55\%.\footnote{The effective
  packing efficiency is less than the maximum close-packed efficiency
  of 74\% due to finite-size effects (Paper I).} The parameter space
runs from 0.00--1.25 in $b$ and 0.52--2.50 in $v$ (the units of $b$
and $v$ are defined in \sect{units}; $v_\mathrm{crit} = 2.06$ m
s$^{-1}$ for this model).  The impact parameter values were chosen to
encompass a range of dynamic interactions from head-on collisions to
glancing distortions.  The lowest value of $v$ is twice the value
corresponding to $v_\infty = 0$ (\cf \eqn{vinfty}; smaller $v$ leads
to strong trajectory deflections).  The largest value of $v$ was
chosen to be a moderately high-speed impact to ensure that the
catastrophic dispersal regime was entered.

Figure \ref{f:modelA} summarizes the results of this model (Figs.\ 
\ref{f:strips}(a)--(c) give snapshots of three distinct outcomes).
The shapes in \fig{modelA} trace the projected silhouettes of the
largest post-encounter fragment at the end of each run.  We have used
nested squares of different line styles to divide our results into
three mass regimes.  A solid-line inner square indicates that the
largest fragment contains 90\% or more of the total mass of the
system, \ie nearly perfect accretion.  A dashed-line inner square
indicates that the largest fragment contains at least 50\% but less
than 90\% of the total mass.  The remaining cases contain less than
50\% of the total mass in the largest fragment, \ie net erosion.  Note
if there is no mass loss or exchange during the encounter the largest
fragment will contain 50\% of the total mass of the system by
definition.  We see in this model that 18 out of 55 runs (33\%) result
in net mass loss, although we caution that several cases are just on
the border of 50\%.

The general trends in \fig{modelA} are twofold, namely, as the
encounter speed increases, the size of the largest fragment decreases,
and as the impact parameter increases, the axis ratio increases, up to
a certain point.  Higher encounter speeds imply larger kinetic energy
so it is more likely for the system to become unbound.  Larger impact
parameters imply larger angular momentum which results in an increase
in the axis ratio until the critical spin value of the combined rubble
pile is reached.  In addition to the general trends, the middle-mass
group has two distinct populations that reflect their formation
history.  The small $b$, large $v$ group (top left in the figure)
represents a net loss of mass of 10\%--50\% from the system.  The
large $b$, small $v$ group (lower right) represents grazing collisions
with little mass loss or exchange.

We note that for the head-on case our definition of $v_\mathrm{crit}$
does not correspond to critical dispersal, rather, critical dispersal
seems to occur at $\sim 1.9v_\mathrm{crit}$.  This probably reflects
the fact that the energy of the collision is not immediately
transported to all of the particles and that the voids in between the
particles decrease the efficiency of energy propagation.  Moreover, we
did not take into account $\epsilon_n$ in the definition of
$v_\mathrm{crit}$.  Regardless, $v_\mathrm{crit}$ is intended as an
approximate scaling only.

More detailed results for this model are given in \tbl{modelA}.  In
the table, $b$ and $v$ have the usual definitions; $M_\mathrm{rem}$ is
the mass fraction of the largest post-encounter remnant; $P$ is its
instantaneous spin period in hours; $\varepsilon$ is the remnant's
``ellipticity'': $\varepsilon \equiv 1 - \frac{1}{2}(q_2 + q_3)$,
where $q_2 \equiv a_2/a_1$, $q_3 \equiv a_3/a_1$, and $a_1 \ge a_2 \ge
a_3$ are the semi-axes ($\varepsilon = 0$ is a sphere);
$f_\mathrm{mix}$ is given by \eqn{mixing}; $M_\mathrm{acc}$,
$M_\mathrm{orb}$, and $M_\mathrm{esc}$ are the mass fractions that are
accreting, orbiting, and escaping from the largest remnant,
respectively;\footnote{To be considered accreting, a clump must have
  $q < r + R$, where $q$ is the close-approach distance to the
  remnant, and $r$ and $R$ are the radii of minimal spheres enclosing
  the clump and remnant, respectively. This differs somewhat from
  Paper I.} and $n_1$, $n_2$, and $n$ are the number of single
particles, two-particle groups, and discrete rubble piles (\ie groups
with three or more particles), respectively, at the end of the run.
The $M_\mathrm{rem}$ column of \tbl{modelA} compliments \fig{modelA}
by providing a finer gradation of the remnant mass.  Note that
$M_\mathrm{rem} + M_\mathrm{acc} + M_\mathrm{orb} + M_\mathrm{esc} =
1$.

\tbl{modelA} shows how the remnant spin $P$ is coupled to the angular
momentum in each run.  Since there are no external torques in the
system, angular momentum is conserved.  In the case of head-on
collisions ($b = 0$), there is exactly zero total angular momentum,
which accounts for the large remnant $P$ values (\ie low spin).  $P$
is never infinite in these cases because some particles escape and
carry angular momentum away from the remnant, even in the slowest
collision case ($v = 0.52$).  At higher collision speeds, more mass is
carried away from the system, generally resulting in smaller $P$
values.  As $b$ increases, so does the net angular momentum, resulting
in faster spins (smaller $P$).  This trend continues until $b \sim 1$
which corresponds to a grazing collision.  In this case, the encounter
generally does not result in a merger so the ``remnant'' is
effectively one of the initial bodies plus or minus some mass
exchange.  Mass exchange and/or tidal torquing following deformation
converts orbital angular momentum into spin angular momentum.  As $b$
increases further, there is little spin-up, since torquing becomes
less effective.  All of these trends can be seen in the table.

Similarly, $\varepsilon$ depends on the total angular momentum of the
system.  Larger angular momentum allows the remnant to support a more
elongated shape as long as most of the system mass ends up in the
remnant ($\sim$ 75\%, from the table).  Consequently there is also a
relationship between $\varepsilon$ and $P$: smaller $P$ values
correspond to larger $\varepsilon$ values, in general.  The smallest
$P$ in the table is 4.1 h with $\varepsilon = 0.26$; the largest
$\varepsilon$ is 0.45 with $P = 4.3$ h.  These values are within the
classical limit for mass retention at the surface:
\begin{equation}
  P_\mathrm{crit} \simeq \frac{1}{1 - \varepsilon}
  \sqrt{\frac{3\pi}{G\rho}},
\end{equation}
where $\rho$ is the bulk density and we have assumed $a_2 = a_3$.  In
this expression $P_\mathrm{crit} = 2.3$ h for a spherical rubble pile
with $\rho = 2$ g cm$^{-3}$, and increases to infinity as $\varepsilon
\rightarrow 1$.

The sixth column in \tbl{modelA}, labeled $f_\mathrm{mix}$, gives the
\emph{mean} percent mixing fraction and standard deviation after 100
repeated measurements (recall the mixing calculation subvolumes are
chosen randomly---\cf \sect{mixing}).  The errors are a small fraction
of the mean except when the mixing fraction itself is small.  In the
head-on case, $f_\mathrm{mix}$ shows a simple trend of generally
increasing with impact energy with a dip at the highest energy
probably due to increased statistical fluctuation (the remnants are
smaller).  For most $b$ values the situation is more complicated
depending on whether the impactors accrete into a single body or
exchange mass while remaining two separate bodies.  For the largest
$b$, little mass is exchanged, so the bodies remain essentially
unmixed.

The next three columns give dynamical information about the remaining
mass of the system, \ie the material not incorporated in the largest
remnant.  Generally most of this mass is escaping from the largest
remnant ($M_\mathrm{esc}$).  Typically only small amounts ($<$ 10\%)
of mass are accreting ($M_\mathrm{acc}$) and/or orbiting
($M_\mathrm{orb}$).  In two cases, however, $M_\mathrm{acc}$ is close
to 50\%; these are instances of near escape that were too
computationally expensive to run until final accretion and represent
the transition from a high- to medium-mass remnant.

The final three columns contain information about the particle
groupings at the end of each run.  The number of free particles
($n_1$) increases dramatically with $v$, but decreases with $b$.  This
trend is also seen in the number of two-particle groups ($n_2$) and
discrete rubble piles ($n$).  Groups can form either from accretion
among the free particles due to gravitational instability or from
being stripped off as a clump during the collision event.  Note that
$n$ is always at least 1 because the remnant is included.

In summary, the outcomes of this model depend in a natural way on the
total angular momentum and the impact energy of the system (both
related to $b$ and $v$).  Larger $b$ results in more elongated
remnants with higher spins and reduced mixing.  Larger $v$ results in
greater mass loss and increased mixing.  In the remaining sections we
explore how these trends are modified for non-identical or spinning
bodies.

\subsubsection{Model B: Equal size, spin} \label{s:modelB}

In Model B we added a spin component to the impactors.  The spin
vectors are oriented perpendicular to the orbital plane (\ie along the
$\pm z$ axis).  The rotation period of the impactors is 6 h, the
median rotation period of Near Earth Asteroids (Bottke \etal 1997).
We investigated three cases: in Model B1 the spins of the impactors
have opposite orientation; in Model B2 and B3 the spins have the same
orientation but the impactors have opposite $y$ offsets (\fig{spins}).
By symmetry, these cases test all the unique $z$ angular momentum
combinations (spin $+$ orbital).  The remaining parameters are
identical to those in Model A.

Figures \ref{f:quad}(a)--(c) summarize the results for Models B1, B2,
and B3, respectively (\fig{strips}(d) is a snapshot sequence of a
Model B1 run).  The general trends seen in \fig{quad} are similar to
those seen in \fig{modelA}.  The head-on cases tend to result in
spherical remnants of decreasing mass with increasing $v$.  The
elongation of the remnants tends to increase with an increase in $b$,
up to a point.  Of the three models note that Model B1 is the most
similar to Model A.  This is because Model B1 has the same amount of
net angular momentum in the system since the spin components of the
impactors cancel.  Model B2 and Model B3, however, have smaller and
larger net angular momentum in the system, respectively, than Model A
or B1.  This is reflected in the number of runs with fast-rotating
and/or elongated remnants (\tbl{Pande}).  Models A and B1 have an
intermediate number of runs with extreme $P$ and/or $\varepsilon$
values compared with Model B2 or B3 (Model C is a special case
discussed in the next section).

Model B1 does differ from Model A in one important respect.  As seen
in \fig{quad}(a), some of the remnants in this model (\eg $b = 0.30$,
$v = 1.10$; $b = 0.60$, $v = 0.61$) have unique asymmetries (\ie
broken eight-fold symmetry).  This is because before the encounter one
of the bodies is spinning prograde while the other body is spinning
retrograde with respect to the orbit.  The prograde rotator has larger
angular momentum with respect to the center of mass of the system than
its retrograde counterpart, consequently, it suffers more mass loss.
This is analogous to the resistance of retrograde rotators to tidal
disruption (Richardson \etal 1998).

To summarize other quantitative results, 24\% of the Model B1 runs
resulted in net erosion, while this value was 29\% for B2, and 40\%
for B3.  The mixing statistics are generally similar to those for
Model A, namely that larger disruption resulted in more mixing.  As
for ejecta statistics, again no more than about 2\% by mass remains in
orbit around the remnant in all cases, while a somewhat larger
percentage is destined to reaccrete (no more than $\sim$ 6\%, except
for a few cases where components of a future contact binary were on
slow-return trajectories).  The distribution of fragments ($n_1$,
$n_2$, and $n$) followed similar trends to those of Model A.

\subsubsection{Model C: Unequal size, no spin} \label{s:modelC}

In Model C we used two different-sized impactors with no initial spin:
one large sphere of 1357 particles and 1 km radius, and one small
sphere of 717 particles and 0.46 km radius, keeping the bulk densities
the same (2 g cm$^{-3}$) and the total number of particles similar to
the previous models.  Hence the larger sphere is ten times the mass of
the smaller sphere and the particles in the two impactors are
different sizes (the smaller body has smaller particles, to ensure
adequate resolution).  We caution that the difference in particle
sizes implies different packing efficiencies (porosities) which may
affect the outcome (\cf \sect{hires}).  Both impactors were
equilibrated using the same process as before.  Note that the
parameter space investigated is different from the previous models,
primarily for better sampling of the tidal regime (large $b$, small
$v$).  For this model, $v_\mathrm{crit} = 2.9$ m s$^{-1}$.  As for the
previous models, the minimum $v$ is twice the value corresponding to
$v_\infty = 0$.

In \fig{quad}(d) it is evident that most collisions result in net
growth of the larger body (only 9 cases, or 16\%, result in net
erosion \ie remnants with less than 90\% of the total mass of the
system).  None of the encounters resulted in critical \emph{dispersal}
and only the highest-speed cases resulted in violent disruption of
the combined system (\ie $b = 0$, $v = 2.5$).  Also the remnants are
all roughly spherical (the largest $\varepsilon = 0.26$).
\fig{strips}(e) shows a typical encounter: the small body is
pulverized and occasionally planes off a chunk of the larger body (so
$n_1$ is typically a few hundred in all runs except the most grazing,
while $n_2$ and $n$ remain small, $\ale\ 10$).  Most of the smaller
fragments escape the system, a tiny fraction ($< 1$\% by mass) go into
orbit around the remnant, while the remaining fragments return and
blanket the remnant in the equatorial plane.  The largest
concentration of smaller particles is at the impact site.  The
rotation periods of the remnants in this model are typically long
compared to those of the previous models due to the larger rotational
inertia of the bigger impactor.  Finally, there was little tidal
interaction seen in any of the cases, suggesting even the minimum $v$
was too large.  Unfortunately, smaller $v$ would result in stronger
path deflections, making interpretation difficult.

\subsection{Coefficient of Restitution Test} \label{s:cofr}

The energy change in the center-of-mass frame of a system of two
smooth, colliding spheres is given by (\eg Araki and Tremaine 1986):
\begin{equation}
  \Delta E = - \frac{1}{2} \mu (1 - \epsilon_n^2) v_n^2,
\end{equation}
where $v_n$ is the component of relative velocity normal to the mutual
surfaces at the point of contact, and $\mu$ and $\epsilon_n$ have the
usual definitions.  Hence as $\epsilon_n \rightarrow 0$, all the
impact energy---less a geometric factor that depends on $b$---is
dissipated.  Although a collision between two rubble piles consists of
many individual particle collisions, the dependence of $\Delta E$ on
$\epsilon_n$ suggests that $Q^\star_D$ will depend on $\epsilon_n$ in
a similar way, namely that smaller $\epsilon_n$ implies larger
$Q^\star_D$.

A simple test bears this out.  \tbl{cofr} and \fig{cofr} summarize the
effect of varying $\epsilon_n$ for one of the Model A runs ($b =
0.15$, $v = 2.00$; \cf \fig{strips}(b)).  The general trend is clear:
as $\epsilon_n$ decreases, the size of the largest remnant increases
(note the large $M_\mathrm{acc}$ value for the $\epsilon_n = 0.2$
case, indicating that a big fragment is about to merge with the
remnant, giving it the largest mass of all the runs).  Runs with
smaller $\epsilon_n$ form discrete rubble piles out of the collision
debris faster and more efficiently than those with larger
$\epsilon_n$.  For $\epsilon_n = 1$, no rubble piles actually form.
The strong dependence on $\epsilon_n$ suggests that further study is
needed to determine the value most representative of true rubble pile
collisions.

\subsection{High-Resolution Models} \label{s:hires}

In order to test the degree to which particle resolution affects the
collision outcome, we performed two high-resolution runs using
parameters drawn from Model A.  Each impactor for this test consisted
of 4,995 identical particles, more than 5 times the number used in the
parameter space runs (\sect{params}).  The progenitor needed 1 CPU day
to equilibrate using 2 processors on the SGI Origin 200.  At
equilibrium, the code was performing $\sim 4 \times 10^4$ collisions
per step, with each step requiring $\sim$ 140 s wallclock time.  The
enhanced packing efficiency of the high-resolution impactors plus
their randomized orientations makes detailed comparison with the
low-resolution runs difficult.  However, we would expect the general
trends to be similar (\ie outcome class, etc.).  Figure
\ref{f:hires}(a) shows snapshots shortly after the initial impact
comparing the low- and high-resolution Model A runs with $b = 0.30$,
$v = 1.25$.  Figure \ref{f:hires}(b) shows post-reaccretion snapshots
for $b = 0.60$, $v = 0.61$.  These runs were chosen because they are
moderately well separated in $b$-$v$ space while still being
representative of the complex intermediate-energy regime (\cf
\fig{modelA}).  The expense of these calculations precluded a more
thorough sampling.

Both high-resolution runs in this test show evolution similar to that
of their low-resolution counterparts.  In \fig{hires}(a), the
impactors mutually penetrate and lose most of their relative orbital
energy (the bodies will eventually accrete into a single massive
remnant).  Note the presence of the ``mass bridge'' between the two
bodies in both cases.  The rotational phase and penetration distance
differ somewhat, perhaps indicating that higher resolution (and hence
lower porosity) gives rise to more efficient dissipation, perhaps by
increasing the degrees of freedom.  In
\fig{hires}(b), the final shape of the reaccreted body at low and high
resolution is similar, but there is more structural detail in the
high-resolution remnant, \eg the depression on the upper surface. The
remnant mass and rotation period are comparable at this instant: 0.985
and 4.2 h respectively at low resolution; 0.987 and 4.1 h at high
resolution.  We conclude that higher resolution may give insight into
the more detailed aspects of reaccumulation, but low resolution is
sufficient for a broad sampling of parameter space.

\section{DISCUSSION} \label{s:discuss}

\subsection{Critical Dispersal Threshold}

Despite the large number of runs carried out for this investigation,
the data are still too sparse in each model to reliably derive a
generalized expression for the retained mass (remnant plus accreting
and orbiting material) as a function of $b$ and $v$, \ie $1 -
M_\mathrm{esc} = f(b,v)$.  However, we can solve for the critical
contour $f(b,v) = 0.5$, which is well sampled by our choice of
parameter space (for Model C, we solve for $f(b,v) = 0.9$, the point
of net erosion for the larger impactor).  Our method is to perform
bi-linear interpolation of our $b$ \vs $v$ results onto a regular
grid, root solve using Newton's method for the $v$ value that gives a
remnant mass of 0.5 at each grid line in $b$ (we chose 20 lines for
smooth sampling), and fit the resulting values to a functional form.
After some experimentation, we found the contour is best represented
by a Gaussian:
\begin{equation}
  v|_{f = 0.5} \equiv v_\star = \alpha \exp \left[ - \frac{(b -
      \beta)^2}{\gamma} \right] + \delta , \label{e:fits}
\end{equation}
where $\alpha$, $\beta$, $\gamma$, and $\delta$ are parameters to be
determined by non-linear least-squares fitting.

\tbl{fits} gives the best-fit values of the Gaussian parameters along
with their 1-$\sigma$ uncertainties for each of the parameter space
models.  Note that the fits are marginally consistent with $\beta =
0$, \ie no $b$ offset, except for Model C.  The differences between
the fits (except Model C) are slight, but they follow the trend
mentioned in \sect{modelB}, namely that Model B2 has a higher
disruption threshold than Model B3, with Models A and B1 having
intermediate thresholds.  Model C has a broader distribution that is
somewhat offset in $b$, but inspection of the other models shows
similar trends for the 0.9 contour.

\subsection{Debris Size Distributions}

Combining all 275 runs of Models A, B, and C, we find the largest
primary (remnant) mass is 0.999 (so there were no perfect mergers),
the largest secondary mass is 0.498 (there was always some grazing
mass exchange, at least in Models A and B), and the largest tertiary
mass is 0.073.  The smallest primary mass is 0.038.  In Models A and
B, the most common outcome was an even split in mass between the
primary and secondary, since most runs at moderate to large $b$
resulted in little to no mass exchange between the impactors.  For $b
\le 0.30$ ($\phi \le 17^\circ$), the normalized primary mass function
is well approximated by a curve of the form $n(m) \propto 1/(1 -
m^2)$, $m < 1$ (\fig{mfunc}).

\subsection{Debris Spatial Distributions} \label{s:iso}

In cases where debris escapes the central remnant, the ejected
material is invariably concentrated in a plane normal to the orbital
($z = 0$) plane, although in some cases material can be spread out in
the orbital plane as two returning fragments coalesce.  For our
head-on collisions ($b = 0$), the dispersal plane is normal to the $x$
axis.  For $b > 0$, the plane is initially normal to the impact angle
$\phi$, but rotational inertia from the orbital motion causes the
dispersal plane to overshoot this value.  As usual, initial spin may
help or hinder this process (note for Model B3, $\phi \sim
-\sin^{-1}b$).

Figure \ref{f:iso} illustrates the anisotropic distribution of ejecta
as projected to the orbital plane for 4 of the 5 runs shown in
\fig{strips}.  In the equal-mass cases the angular distribution is
bi-modal, with the peaks roughly 180$^\circ$ apart.  For the plot
representing \fig{strips}(b), the peaks are of unequal amplitude since
the remnant is relatively small and displaced from the system center
of mass.  The one unequal-mass case is uni-modal, indicating that
debris was scattered preferentially in one direction (roughly
30$^\circ$ measured counterclockwise from the $x$ axis), as seen in
\fig{strips}(e).

Generally the $z$ distributions are sharply peaked near the largest
remnant but some particles end up many hundreds of km away.  Recall
that the tidal field of the Sun is not included in our simulations.
If it were, these particles would be well outside the remnant's Hill
sphere at 1 AU (for example): \newlength{\mystrut}
\settoheight{\mystrut}{R}
\begin{equation}
  r_H = (\mathrm{210\ km}) \left[
    \frac{a\rule{0pt}{\mystrut}}{\mathrm{1\ AU}} \right] \left[
    \frac{R}{\mathrm{1\ km}} \right] \left[ \frac{\rho}{\mathrm{2\ g\ 
        cm}^{-3}} \right]^{1/3}.
\end{equation}
Regardless, most of these particles would escape the remnant, even
without the solar tides.

\subsection{Outcome Probability} \label{s:outcome}

For a given impact parameter and speed distribution, the probability
of a net accretional (as opposed to net erosional) outcome can be
estimated from \eqn{fits}.  Suppose we set $b = 0.7$, which
corresponds to $\phi = 45^\circ$, the most probable impact angle for
randomly flying projectiles striking a spherical target (Love and
Ahrens 1996).  A monodispersive population of bodies with a Maxwellian
distribution of speeds whose rms equals the escape speed $v_e$ from a
particle's surface has the following normalized distribution function
in relative speed (\eg Binney and Tremaine 1987, Problem 7-3):
\begin{equation}
  g(v)dv = \frac{1}{2\sqrt{\pi}v_e^3} \exp \left( - \frac{v^2}{4v_e^2}
  \right) v^2 dv.
\end{equation}
The probability of a net accretional impact for hyperbolic encounters
with $b = 0.7$ is then:
\begin{equation}
  P[f(b = 0.7,v) \ge 0.5] = \frac{\int_{v_\circ}^{v_\star}
    g(v)dv}{\int_{v_\circ}^\infty g(v)dv}, \label{e:prob}
\end{equation}
where $v_\circ$ is the initial speed corresponding to $v_\infty = 0$
from \eqn{vinfty} and $v_\star$ is obtained from \eqn{fits}.  For
Model A, we have $v_e = 0.51$, $v_\circ = 0.22$, and $v_\star = 0.81$.
Solving \eqn{prob} numerically we find the probability of an
accretional impact in this case is 26\%.  The probability of erosion
is $1 - P(f \ge 0.5) = 74\%$.

The full accretional cross section is obtained by integrating
\eqn{prob} over all impact parameters.  The net accretion probability
is then the ratio of this value to the geometrical cross section:
\begin{equation}
  P[f(b,v) \ge 0.5] = \frac{1}{\pi \int_0^1 bdb} \pi\int_0^1 bdb
  \frac{\int_{v_\circ(b)}^{v_\star(b)}
    g(v)dv}{\int_{v_\circ(b)}^\infty g(v)dv},
\end{equation}
where the dependence of $v_\circ$ and $v_\star$ on $b$ has been made
explicit.  Solving this equation we find the accretion probability
increases to 36\% only, since head-on collisions are relatively rare.
Such a low value implies that this population of rubble piles would
not go on to form planets but would instead grind itself down to dust.
If rubble piles were common during the early stages of planet
formation then perhaps collisions were more dissipative than modeled
here.  Alternatively, the accretion probability may be enhanced when
there is a distribution of masses, a possibility that can only be
tested with more simulations using impactors of varying size.

\subsection{Comparison with Previous Work (Gravity Regime)}

Using the fits to \eqn{fits} we can estimate the value of $Q^\star_D$
(recall this is for $\epsilon_n = 0.8$; further runs are needed to
determine the dependence on dissipation).  Restricting ourselves to
Model A with $b = 0$, we find $Q^\star_D \sim 1.9$ J kg$^{-1}$.  This
lies very close to the Holsapple (1994) and gravitational binding
energy curves described in Love and Ahrens (1996; see in particular
their Eq.\ (2) and Fig.\ 7).  It lies well off the extrapolation of
their SPH results.  In their paper they suggest that the discrepancy
between their results and analytic or experimental results may arise
from: 1) the local rather than global deposition of impact energy at
the surface of the target; 2) the difference between the role of
gravity in self-compression and ejecta retention; 3) the finite size
of the projectile.  The present work however is similar to Love and
Ahrens (1996) in all these respects, which suggests the difference may
be attributable solely to the adopted equation of state (an
incompressible fluid in our case, compared with the Tillotson equation
of state for granite in theirs).  Note that the SPH curve plotted in
Fig.\ 7 of Love and Ahrens (1996) was for an impact angle of $\phi =
45^\circ$, but this would amount to less than an order of magnitude
difference in $Q^\star_D$.

If the outcome truly depends solely on the gravitational binding
energy (ignoring the effect of dissipation for now as this requires
further study), then we would expect $Q^\star_D \propto M/R \propto
R^2 \propto M^{2/3}$.  From our Model A point we can estimate the
constant of proportionality: $Q^\star_D \sim 1.2 \times 10^{-6} R^2
\sim 2.9 \times 10^{-9} M^{2/3}$.  Further models with different $M$
are needed to confirm this result (our Model C case failed to sample
the critical dispersal regime, so we cannot use it here).

Watanabe and Miyama (1992) found $M_\mathrm{esc} \propto v^3$ for
their low-speed, head-on SPH models (see Eq.\ (3.5.1) in their paper).
We find a similar trend.  For the $b = 0$ outcomes of Model A, a
least-squares fit to the form
\begin{equation}
  M_\mathrm{esc} = \alpha v^\beta
\end{equation}
yields $\alpha = 0.06 \pm 0.02$ and $\beta = 3.2 \pm 0.1$.  Evidently
this relation must break down at large $v$, otherwise $M_\mathrm{esc}$
would exceed unity.  Indeed our only significant outlier is for our
highest $v$ value (2.50), with $M_\mathrm{esc}$ in this case $\sim$
20\% below the curve.

\section{CONCLUSIONS} \label{s:concl}

In summary, we have conducted a series of numerical simulations to
create a partial map of the parameter space of rubble pile collisions
at low impact speeds.  The general trends can be summarized as
follows: 1) larger impact angles result in more elongated,
faster-spinning remnants; 2) larger impact speeds result in greater
mass loss and increased mixing of the remnant; 3) initial impactor
spin can increase or reduce the rotation period and elongation of the
remnant.  It is also possible to create asymmetric shapes if the
impactors have oppositely oriented spins.  These general trends are
directly related to the total energy and angular momentum of the
system.  In cases where one impactor is significantly larger than the
other (Model C), the smaller body generally disrupts completely on
impact, sometimes removing a modest fraction of the surface of the
target body and sometimes redepositing material along the remnant's $z
= 0$ equator.

We have been able to generate a wide variety of remnant shapes,
including spheroids, ellipsoids, contact binaries (peanut shaped and S
shaped), and shapes with broken eight-fold symmetry.  It proved
considerably difficult to get a significant amount of material to
orbit the remnant; most debris (98\%) either accreted onto the remnant
or escaped from the system.  We found no detached binaries of
significant size, but $\sim$ 10\% of the remnants in Model A and B are
contact binaries.  The coefficient of restitution appears to play a
more important role in collisions than in tidal disruption and can
strongly affect the number and size of post-impact rubble-pile
fragments.  Increased particle resolution (or reduced porosity)
appears to augment dissipation and give rise to more complex shapes,
but the effects are modest over a factor of 5 in particle number.

We found that the impact speed needed for critical dispersal is well
represented by a Gaussian function of impact parameter.  Given a
velocity distribution it is possible to estimate the probability of
either impactor gaining or losing mass as a result of the collision.
At low impact angles with equal-size impactors the remnant mass
function is roughly proportional to $1/(1 - m^2)$.  Secondary and
tertiary masses are typically finite but small, except for
near-grazing encounters.  Most material is ejected anisotropically in
a plane perpendicular to the axis of the initial motion and in some
cases the debris can coalesce into smaller rubble piles.

We found $Q^\star_D \sim 2$ J kg$^{-1}$ for the head-on collisions in
Model A and that $M_\mathrm{esc} \propto v^{3.2}$.  The former result
is in rough agreement with the theoretical gravity-regime model of
Holsapple (1994).  The latter relation agrees with Watanabe and Miyama
(1992).  We find that km-sized rubble piles in general are much easier
to disperse than extrapolation of the strengthless granite models of
Love and Ahrens (1996) would suggest.  This may be due in part to our
conservative choice for $\epsilon_n$ (0.8).  Although more work needs
to be done, we believe our simulations may provide a numerical basis
for parameterizing collisions during the early stage of planet
formation, when the planetesimals are dynamically cool and the
dominant sizes are still $\ale$ 10 km.

\subsection{Future Work} \label{s:future}

In this study we were restricted to investigating the dependence of
collision outcome primarily on impact parameter and impact speed.  We
also examined a few spin combinations, a model with unequal masses,
and a single run with various values of the restitution coefficient.
But the parameter space is truly vast.  Naturally we would like to
test more values for the parameters we have already investigated,
particularly the coefficient of restitution and the dependence on
porosity.  We also need a finer grid at small speed and near-grazing
separation to fully investigate the tidal regime (this would also
provide better data for comparison with stellar system collision
models, \eg Davies \etal 1991).  However there are many other new
parameters to explore.  We would like to test the effect of changing
the spin-axis orientations (beyond pure prograde or retrograde).  We
suspect this would result in even more unusual shapes.  Non-spherical
impactors with a variety of sizes would improve realism and provide
better estimates of $Q^\star_D$.  A spectrum of particle sizes could
alter the effective dissipation as smaller particles fill the voids
between larger ones.  Adding surface friction could lead to steeper
slopes and the possibility of simulating crater formation in large
targets.  We plan to add a simple model for compaction to allow higher
impact speeds and compare with the porous models of Housen and
Holsapple (1999).  We would like to track the movement of particles
near the cores of our rubble piles and compare with the surface
particles to study ``scrambling'' in a single rubble pile.  There are
so many possibilities that likely the only practical approach would be
to randomly sample points in this vast parameter space to get a feel
for the overall trends and then concentrate on the most interesting
aspects in detail.  We will carry out such work in the future.

\newpage
\section*{Acknowledgments}

The authors wish to thank the following individuals for their
assistance with this research: E. Asphaug, W. Bottke, C. Dominik, K.
Holsapple, G. Lake, C. Reschke, J. Stadel, B. Titus, and F. van den
Bosch.  This work was supported in part by the NASA HPCC-ESS and Intel
Technology 2000 Programs and a NASA Innovative Research grant.
Ray-traced images were rendered using the Persistence of Vision
Raytracer (POV-Ray version 3.02).

\newpage
\section*{References}
\begin{description}
\item Anderson, R. J. 1993. Computer science problems in astrophysical
  simulation. In \textit{Silver Jubilee Workshop on Computing and
    Intelligent Systems}, pp.\ 48--61. Tata McGraw-Hill, New Delhi.
\item Anderson, R. J. 1996. Tree data structures for $N$-body
  simulation. \textit{37th Symp.\ Foundations of Comp.\ Sci.},
  224--233.
\item \paper {Araki, S., and S. Tremaine} 1986 {The dynamics of dense
    particle disks} Icarus 65 83--109
\item \paper {Asphaug, E., S. J. Ostro, R. S. Hudson, D. J. Scheeres,
    and W. Benz} 1998 {Disruption of kilometer-sized asteroids by
    energetic collisions} Nature 393 437--440
\item \paper {Beaug\'e, C., and S. J. Aarseth} 1990 {$N$-body
    simulations of planetary formation} {Mon.\ Not.\ R. Astron.\ Soc.}
  245 30--39
\item \paper {Bentley, J. L., and J. H. Friedman} 1979 {Data
    structures for range searching} {Computing Surveys} 11 397--409
\item Binney, J., and S. Tremaine 1987. \textit{Galactic Dynamics}.
  Princeton Univ.\ Press, Princeton, NJ.
\item \paper {Bottke, W. F., and H. J. Melosh} 1996a {The formation of
    asteroid satellites and doublet craters by planetary tidal forces}
  Nature 381 51--53
\item \paper {Bottke, W. F., and H. J. Melosh} 1996b {The formation of
    binary asteroids and doublet craters} Icarus 124 372--391
\item Bottke, W. F., Jr., D. C. Richardson, and S. G. Love 1997. Can
  tidal disruption of asteroids make crater chains on the Earth and
  Moon? \textit{Icarus} \textbf{126}, 470--474.
\item Bottke, W. F., Jr., D. C. Richardson, P. Michel, and S. G. Love
  1999. 1620 Geographos and 433 Eros: Shaped by planetary tides?
  \textit{Astron.\ J.} \textbf{117}, 1921--1928.
\item \paper {Davies, M. B., W. Benz, and J. G. Hills} 1991 {Stellar
    encounters involving red giants in globular cluster cores}
  {Astrophys.\ J.} 381 449--461
\item \paper {Durda, D. D.} 1996 {The formation of asteroidal
    satellites in catastrophic collisions} Icarus 120 212--219
\item \paper {Durda, D. D., R. Greenberg, and R. Jedicke} 1998
  {Collisional models and scaling laws: A new interpretation of the
    shape of the Main-Belt asteroid size distribution} Icarus 135
  431--440
\item \paper {Greenberg, R., J. Wacker, C. R. Chapman, and W. K.
    Hartman} 1978 {Planetesimals to planets: Numerical simulation of
    collisional evolution} Icarus 35 1--26
\item \paper {Harris, A. W.} 1996 {The rotation rates of very small
    asteroids: Evidence for ``rubble-pile'' structure} {Proc.\ Lunar
    Planet.\ Sci.\ Conf.} 27 493--494
\item \paper {Holsapple, K. A.} 1994 {Catastrophic disruptions and
    cratering of Solar System bodies: A review and new results}
  {Planet.\ Space Sci.} 42 1067--1078
\item \paper {Housen, K. R., and K. A. Holsapple} 1999 {Impact
    cratering on porous low-density bodies} {Proc.\ Lunar Planet.\ 
    Sci.\ Conf.} 30 1228
\item \paper {Housen, K. R., R. M. Schmidt, and K. A. Holsapple} 1991
  {Laboratory simulations of large scale fragmentation events} Icarus
  94 180--190
\item \paper {Love, S. G., and T. J. Ahrens} 1996 {Catastrophic
    impacts on gravity dominated asteroids} Icarus 124 141--155
\item \paper {Ostro, S. J., and 19 colleagues} 1999 {Radar and optical
    observations of Asteroid 1998 KY26} Science 285 557--559
\item \paper {Petit, J.-M., and M. H\'enon} 1987 {A numerical
    simulation of planetary rings. I---Binary encounters} {Astron.\ 
    Astrophys.} 173 389--404
\item \paper {Richardson, D. C.} 1994 {Tree code simulations of
    planetary rings} {Mon.\ Not.\ R. Astron.\ Soc.} 269 493--511
\item \paper {Richardson, D. C., W. F. Bottke, Jr., and S. G. Love}
  1998 {Tidal distortion and disruption of Earth-crossing asteroids
    (Paper I)} Icarus 134 47--76
\item \inpress {Richardson, D. C., T. Quinn, J. Stadel, and G. Lake}
  1999 {Direct large-scale $N$-body simulations of planetesimal
    dynamics (Paper II)} Icarus
\item \paper {Ryan, E. V., W. K. Hartmann, and D. R. Davis} 1991
  {Impact experiments 3: Catastrophic fragmentation of aggregate
    targets and relation to asteroids} Icarus 94 283--298
\item \paper {Veverka, J, and 16 colleagues} 1997 {Flyby of 253
    Mathilde: Images of a C asteroid} Science 279 2109--2114
\item Wasson, J. T. 1985. \textit{Meteorites---Their Record of Early
    Solar System History}. Freeman, New York.
\item \paper {Watanabe, S., and S. M. Miyama} 1992 {Collision and
    tidal interaction between planetesimals} {Astrophys.\ J.} 391
  318--335
\item \paper {Wetherill, G. W., and G. R. Stewart} 1993 {Formation of
    planetary embryos: Effects of fragmentation, low relative
    velocity, and independent variation of eccentricity and
    inclination} Icarus 106 190--204
\item \paper {Yeomans, D. K., and 12 colleagues} 1997 {Estimating the
    mass of Asteroid 253 Mathilde from tracking data during the NEAR
    flyby} Science 279 2106--2109

\end{description}
%
%
\newpage
\renewcommand{\baselinestretch}{1.0} 
{
  \normalsize 
  \begin{table}
    \centering
    \caption{\bfseries Summary of Model A results (\sect{modelA})}
    \label{t:modelA}
    \begin{tabular}{cccrcr@{\,$\pm$\,}lcccrrr}
      \hline $b$ & $v$ & $M_\mathrm{rem}$ & \multicolumn{1}{c}{$P$} &
      $\varepsilon$ & \multicolumn{2}{@{\,}c@{\,}}{$f_\mathrm{mix}$
        (\%)} & $M_\mathrm{acc}$ & $M_\mathrm{orb}$ & $M_\mathrm{esc}$
      & \multicolumn{1}{c}{$n_1$} & \multicolumn{1}{c}{$n_2$} &
      \multicolumn{1}{c}{$n$} \\ \hline 0.00&0.52&0.992&641.2&0.09&26&4&0.001&0.000&0.007&15&0&1\\
0.00&0.61&0.989&281.7&0.08&29&3&0.001&0.000&0.010&21&0&1\\
0.00&0.75&0.971&220.8&0.06&33&4&0.004&0.000&0.025&55&0&1\\
0.00&0.90&0.936&181.6&0.05&38&5&0.007&0.000&0.057&114&4&1\\
0.00&1.00&0.928&110.8&0.05&50&4&0.007&0.000&0.065&134&2&1\\
0.00&1.10&0.903&114.3&0.04&52&4&0.007&0.000&0.090&182&2&1\\
0.00&1.25&0.843&116.2&0.08&64&4&0.013&0.000&0.145&285&4&3\\
0.00&1.50&0.699&20.7&0.07&73&3&0.022&0.000&0.279&531&11&7\\
0.00&2.00&0.374&19.1&0.04&80&4&0.046&0.016&0.564&938&39&27\\
0.00&2.50&0.098&5.1&0.31&71&4&0.004&0.007&0.891&1328&34&38\\
\hline
0.15&0.52&0.992&11.9&0.07&25&3&0.002&0.000&0.006&16&0&1\\
0.15&0.61&0.984&11.0&0.14&28&3&0.002&0.000&0.014&30&0&1\\
0.15&0.75&0.965&9.1&0.12&30&4&0.005&0.000&0.030&66&0&1\\
0.15&0.90&0.944&7.4&0.14&37&4&0.006&0.000&0.050&105&1&1\\
0.15&1.00&0.924&6.9&0.12&40&4&0.004&0.000&0.072&142&2&1\\
0.15&1.10&0.885&6.1&0.13&47&4&0.016&0.000&0.099&197&4&4\\
0.15&1.25&0.818&5.7&0.07&59&3&0.012&0.001&0.169&324&8&2\\
0.15&1.50&0.695&5.6&0.09&59&3&0.017&0.008&0.280&529&12&6\\
0.15&2.00&0.275&8.4&0.04&52&5&0.009&0.005&0.710&939&24&30\\
\hline
0.30&0.52&0.994&6.9&0.20&21&3&0.001&0.000&0.005&11&0&1\\
0.30&0.61&0.988&6.0&0.20&28&3&0.002&0.000&0.010&23&0&1\\
0.30&0.75&0.974&5.3&0.22&34&3&0.002&0.000&0.025&50&0&1\\
0.30&0.90&0.946&4.6&0.20&37&3&0.006&0.001&0.047&103&0&1\\
0.30&1.00&0.901&4.4&0.24&37&4&0.009&0.010&0.081&184&3&1\\
0.30&1.10&0.887&4.5&0.41&40&4&0.018&0.010&0.085&202&5&2\\
0.30&1.25&0.786&4.5&0.42&42&4&0.013&0.020&0.182&366&10&7\\
0.30&1.50&0.408&7.6&0.09&32&4&0.029&0.007&0.555&481&17&12\\
\hline
0.45&0.52&0.995&5.2&0.33&23&3&0.000&0.000&0.005&10&0&1\\
0.45&0.61&0.986&4.8&0.39&24&4&0.002&0.000&0.012&26&0&1\\
0.45&0.75&0.982&4.3&0.40&26&4&0.002&0.001&0.015&35&0&1\\
0.45&0.90&0.490&9.2&0.13&25&4&0.457&0.009&0.043&105&0&2\\
0.45&1.00&0.469&7.9&0.15&25&4&0.003&0.006&0.523&140&2&2\\
0.45&1.10&0.442&9.6&0.11&23&4&0.008&0.006&0.545&191&4&4\\
0.45&1.25&0.416&9.2&0.06&20&3&0.008&0.004&0.572&295&6&7\\
\hline
0.60&0.52&0.997&4.5&0.40&21&3&0.000&0.000&0.003&6&0&1\\
0.60&0.61&0.989&4.2&0.38&24&3&0.002&0.003&0.007&19&1&1\\
0.60&0.75&0.512&8.1&0.09&20&4&0.468&0.003&0.018&41&0&2\\
0.60&0.90&0.499&8.1&0.13&18&3&0.005&0.005&0.491&93&3&2\\
0.60&1.00&0.460&12.1&0.11&14&2&0.003&0.003&0.535&134&3&4\\
0.60&1.10&0.454&10.9&0.05&11&2&0.013&0.003&0.530&150&5&6\\
\hline
0.75&0.52&0.998&4.1&0.26&23&4&0.000&0.001&0.001&4&0&1\\
0.75&0.61&0.996&4.9&0.39&23&3&0.002&0.001&0.001&8&0&1\\
0.75&0.75&0.493&10.1&0.12&9&2&0.001&0.001&0.506&32&0&2\\
0.75&0.90&0.484&10.0&0.14&7&2&0.003&0.004&0.510&68&2&2\\
0.75&1.00&0.470&20.0&0.13&6&2&0.003&0.002&0.525&111&3&3\\
\hline
0.90&0.52&0.999&4.3&0.45&16&3&0.000&0.001&0.000&2&0&1\\
0.90&0.61&0.504&10.0&0.14&8&2&0.000&0.002&0.494&10&1&2\\
0.90&0.75&0.496&13.9&0.16&4&1&0.002&0.000&0.502&23&0&3\\
0.90&0.90&0.492&14.5&0.12&3&1&0.001&0.000&0.507&42&0&4\\
\hline
1.00&0.52&0.502&9.6&0.12&6&2&0.000&0.001&0.498&5&0&2\\
1.00&0.61&0.502&13.8&0.11&5&1&0.002&0.001&0.495&13&0&2\\
1.00&0.75&0.499&16.6&0.20&3&1&0.001&0.000&0.499&16&1&2\\
\hline
1.10&0.52&0.501&12.6&0.09&4&1&0.000&0.001&0.498&4&0&2\\
1.10&0.61&0.499&17.2&0.12&2&1&0.001&0.000&0.500&5&0&2\\
\hline
1.25&0.52&0.501&48.4&0.07&1&1&0.000&0.000&0.499&2&0&2\\
 \hline
    \end{tabular}
  \end{table}
  } 
\renewcommand{\baselinestretch}{1.5}
\clearpage
\begin{table}
  \centering
  \caption{\bfseries Comparison of runs with extreme $\bsym{P}$ and
    $\bsym{e}$ values (\sect{modelB})}
  \label{t:Pande}
  \bigskip
  \begin{tabular}{ccc}
    \hline
    Model & No.\ with $P \le 5$ h & No.\ with $e \ge 0.35$ \\
    \hline
    A & 11 & 8 \\
    B1 & 10 & 8 \\
    B2 & 10 & 2 \\
    B3 & 15 & 10 \\
    C & 0 & 0 \\
    \hline
  \end{tabular}
\end{table}
\clearpage
\begin{table}
  \centering
  \caption{\bfseries Effect of varying dissipation in Model A run
    $\bsym{b}$ = 0.15, $\bsym{v}$ = 2.00 (\sect{cofr})}
  \label{t:cofr}
  \bigskip
  \begin{tabular}{cccccrrr}
    \hline
    $\epsilon_n$ & $M_\mathrm{rem}$ & $M_\mathrm{acc}$ &
    $M_\mathrm{orb}$ & $M_\mathrm{esc}$ & $n_1$ & $n_2$ & $n$ \\
    \hline
    0.2 & 0.196 & 0.177 & 0.028 & 0.598 &  420 & 29 & 44 \\
    0.5 & 0.364 & 0.006 & 0.004 & 0.626 &  483 & 31 & 33 \\
    0.6 & 0.301 & 0.007 & 0.007 & 0.685 &  597 & 31 & 41 \\
    0.7 & 0.284 & 0.029 & 0.006 & 0.682 &  746 & 32 & 39 \\
    0.8 & 0.275 & 0.009 & 0.005 & 0.710 &  939 & 24 & 30 \\ 
    0.9 & 0.040 & 0.006 & 0.005 & 0.949 & 1484 & 63 & 26 \\
    1.0 & 0.001 & 0.000 & 0.000 & 0.999 & 1904 &  3 & 0  \\
    \hline
  \end{tabular}
\end{table}
\clearpage
\begin{table}
  \centering
  \caption{\bfseries Parameter fits to \eqn{fits} for
    $\bsym{M_\mathrm{rem} = 0.5}$ contours, $\bsym{b \le 1}$}
  \label{t:fits}
  \bigskip
  \begin{tabular}{ccr@{\,$\pm$\,}lcc}
    \hline
    Model & $\alpha$ & \multicolumn{2}{c}{$\beta$} & $\gamma$ & $\delta$ \\
    \hline
    A  & $1.10 \pm 0.03$ &  0.02  & 0.01  & $0.17  \pm 0.01$  & $0.74  \pm 0.02$  \\
    B1 & $0.98 \pm 0.01$ & -0.024 & 0.005 & $0.190 \pm 0.005$ & $0.846 \pm 0.006$ \\
    B2 & $1.2  \pm 0.1$  & -0.02  & 0.04  & $0.27  \pm 0.05$  & $0.67  \pm 0.06$  \\
    B3 & $0.89 \pm 0.03$ &  0.00  & 0.02  & $0.17  \pm 0.02$  & $0.70  \pm 0.02$  \\
    C$^\dag$
       & $0.10 \pm 0.03$ &  0.12  & 0.04  & $0.010 \pm 0.009$ & $1.25  \pm 0.02$  \\
    \hline
    \multicolumn{6}{l}{$^{\dag}M_\mathrm{rem} = 0.9$ contour, $b \le 0.5$}
  \end{tabular}
\end{table}
\clearpage
\section*{Figure Captions}
\begin{description}

  \figcap{strips} Snapshots of rubble pile collisions from
  representative runs as seen in the center-of-mass frame. The models
  and runs are: (a) Model A, $b = 0.00$, $v = 1.00$; (b) Model A, $b =
  0.15$, $v = 2.00$; (c) Model A, $b = 0.90$, $v = 0.52$; (d) Model
  B1, $b = 0.30$, $v = 1.10$; and (e) Model C, $b = 0.50$, $v = 1.25$.
  The arrow of time is to the right. The interval between frames is
  not regular: the snapshots were chosen to highlight distinct stages
  in the evolution of each run. In run (b), the final two frames have
  been brightened for clarity.

  \figcap{modelA} Projected shape of the largest remnant at the end of
  each Model A run as a function of $b$ and $v$. At this scale each
  grid square measures 4 km on a side. Solid inner squares indicate
  remnants that retain at least 90\% of the system mass; dashed
  squares indicate remnants with at least 50\%. Critical dispersal
  generally corresponds to the transition from solid to dashed,
  although in some cases a sizeable fragment may be about to accrete
  with the remnant. \tbl{modelA} gives additional data for this model.

  \figcap{spins} Illustration of the spin sense for the Model B
  impactors. In Model B1, the impactors have opposite spin; in B2 they
  have the same spin, oppositely aligned with the orbital angular
  momentum; in B3 the spins are aligned with the orbital angular
  momentum.

  \figcap{quad} Remnant shapes for the remaining parameter space
  models. The model is indicated in the top left of each plot. Compare
  with \fig{modelA}.

  \figcap{cofr} Snapshots showing the effect of varying $\epsilon_n$
  for the Model A run with $b = 0.15$, $v = 2.00$ (\cf
  \fig{strips}(b)). Each snapshot was taken about 6.5 h after impact.
  The chosen $\epsilon_n$ value and camera zoom-out factor are shown
  in the top left of each frame. For clarity, no color or shading
  distinction is made between the particles of the original impactors,
  and the $\epsilon_n = 0.9$ and $\epsilon_n = 1.0$ frames have been
  brightened. From these snapshots and the statistics in \tbl{cofr} it
  can be seen that rubble pile formation favors smaller $\epsilon_n$
  values. The differences at the extremes are dramatic.

  \figcap{hires} Comparison of two Model A runs performed at low
  resolution (955 particles per rubble pile; left column) and high
  resolution (4,995 particles per rubble pile; right column). The run
  parameters are: (a) $b = 0.30$, $v = 1.25$; (b) $b = 0.60$, $v =
  0.61$. The evolution is similar in both cases, with differences
  attributable to packing efficiency, initial orientation, and
  possibly enhanced dissipation at higher resolution.

  \figcap{mfunc} Primary (solid line), secondary (dotted), and
  tertiary (short dashed) mass fractions of every Model A and B run
  with $b \le 0.30$, sorted by primary mass.  The long-dashed line was
  obtained by integrating a rough fit to the primary mass function of
  the form $1/(1 - m^2)$.

  \figcap{iso} Debris dispersal patterns in the initial orbital plane
  relative to the largest remnant for the runs labeled (a), (b), (d),
  and (e) in \fig{strips}.  The $\theta$ histograms are binned in
  5$^\circ$ increments.  Only particles with projected distances in
  the $z$ plane exceeding twice the remnant radius were included.

\end{description}
%
%
\putfig{strips}{\textwidth}{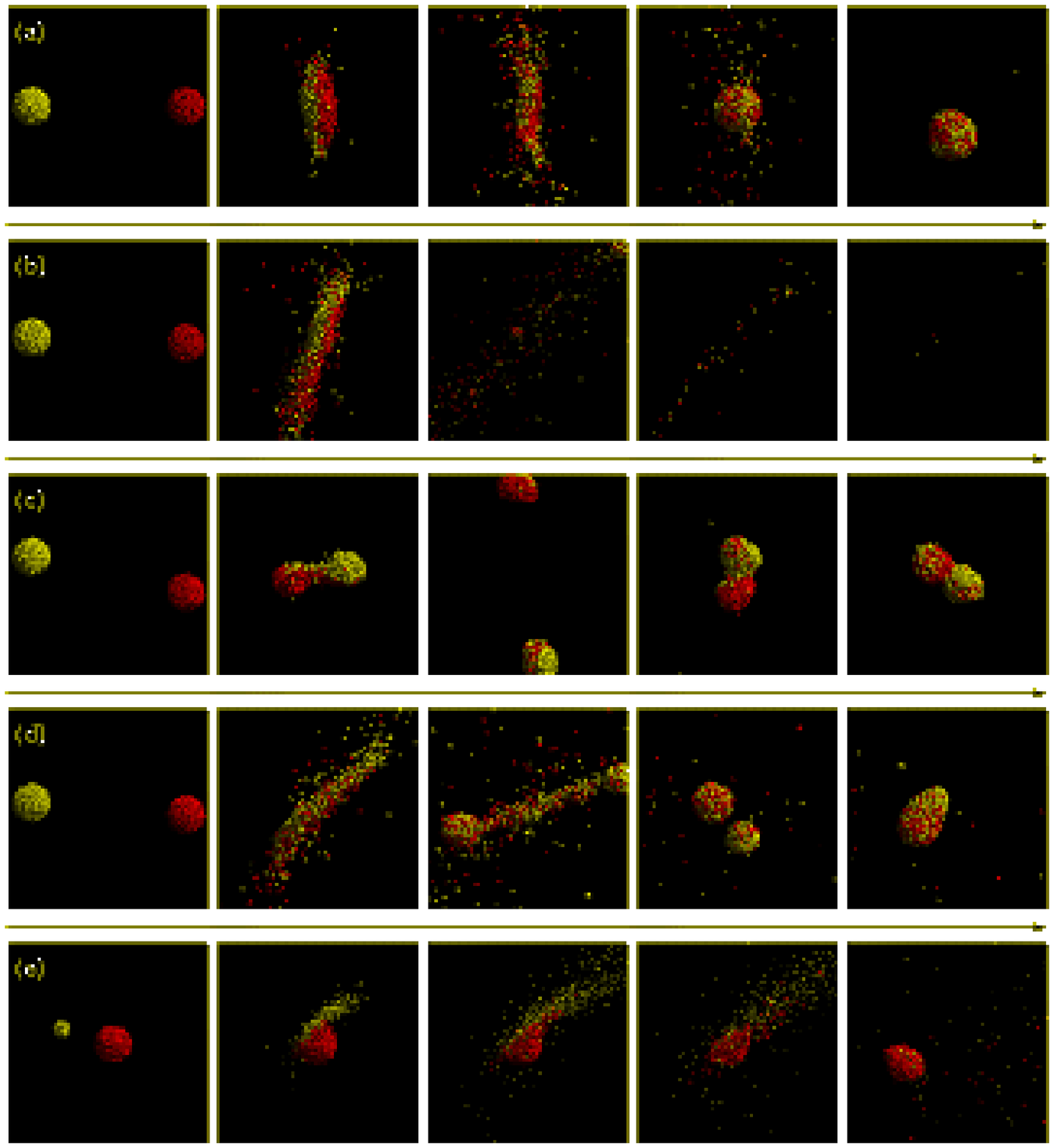}
\putfig{modelA}{\textwidth}{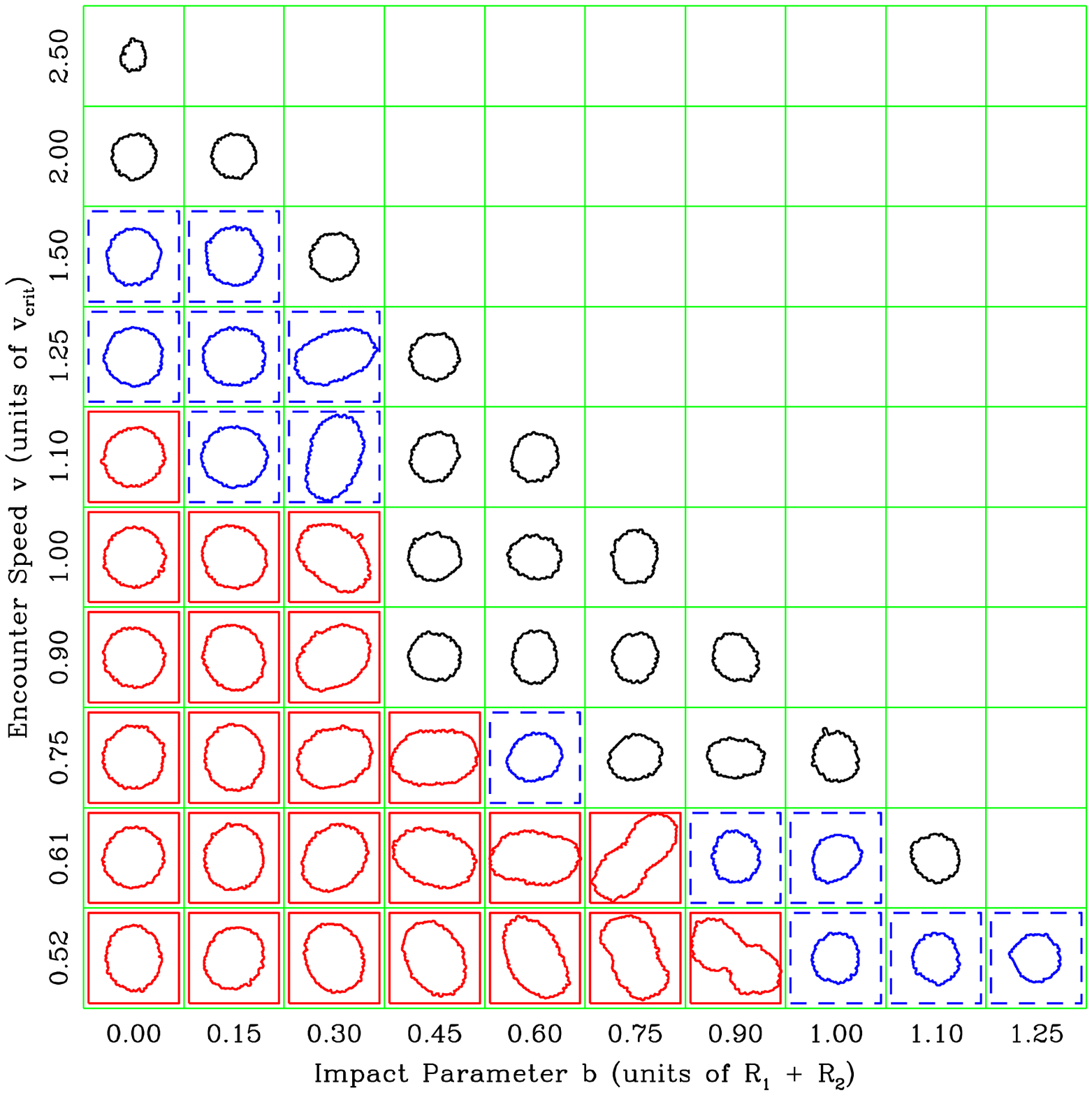}
\putfig{spins}{5in}{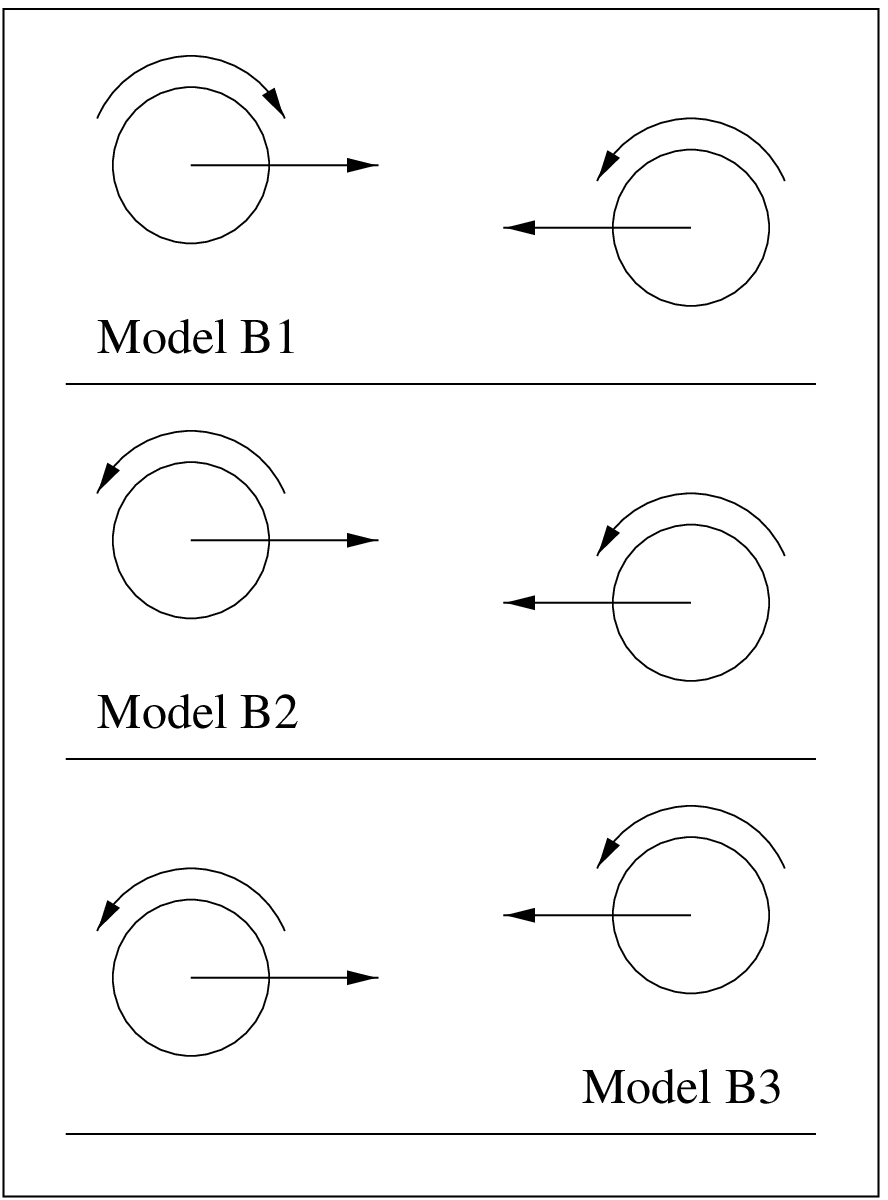}
\putfig{quad}{\textwidth}{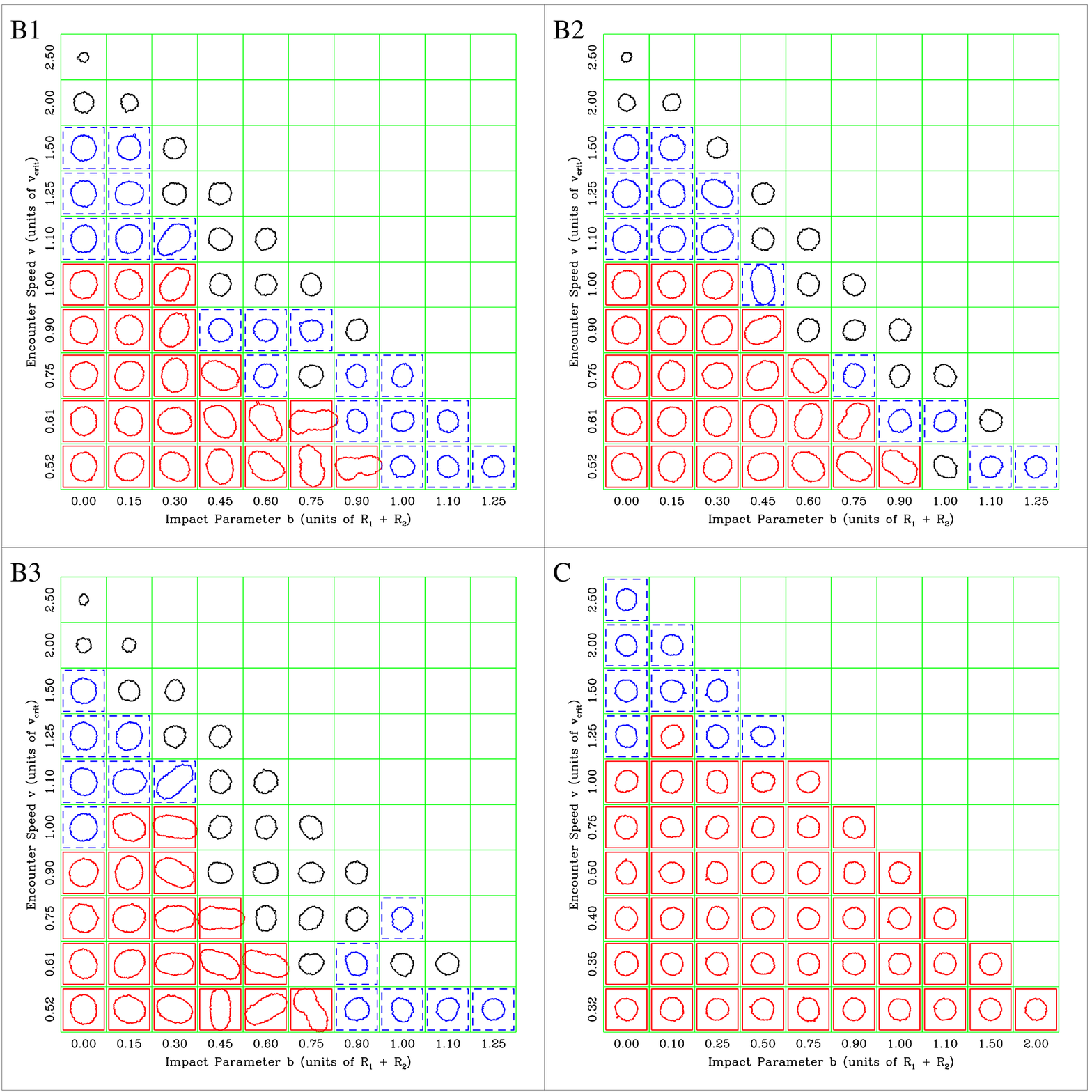}
\putfig{cofr}{\textwidth}{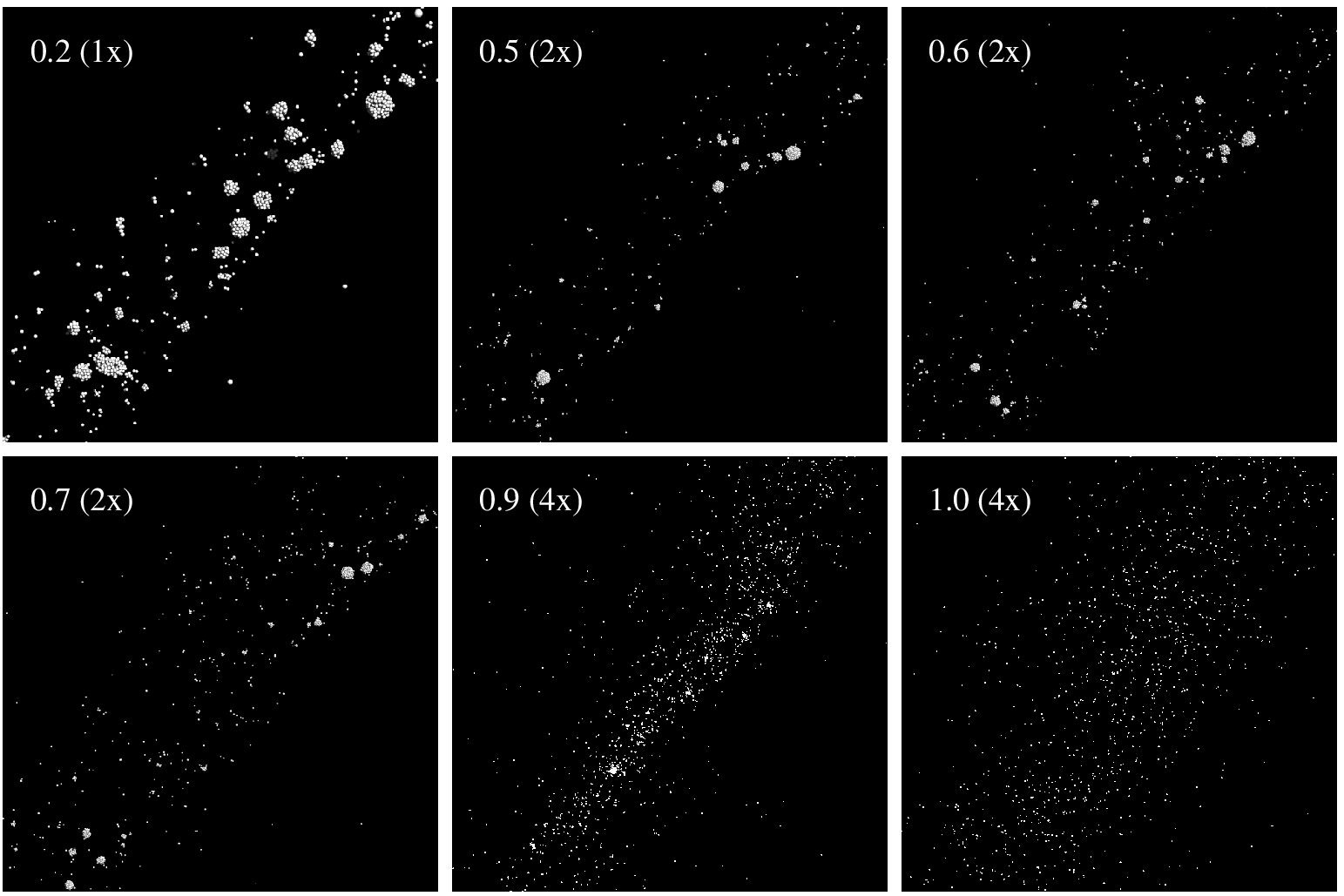}
\putfig{hires}{\textwidth}{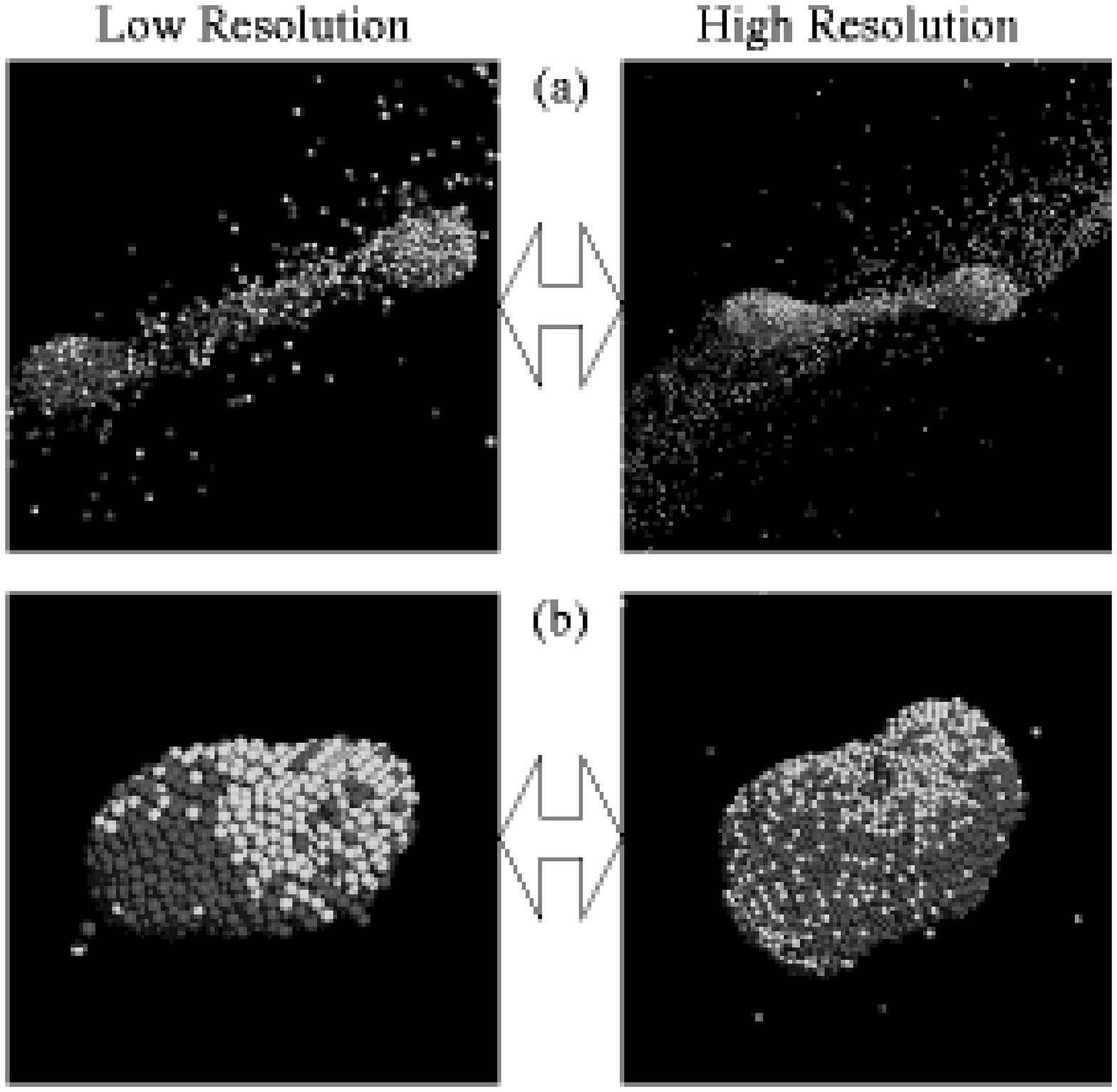}
\putfig{mfunc}{\textwidth}{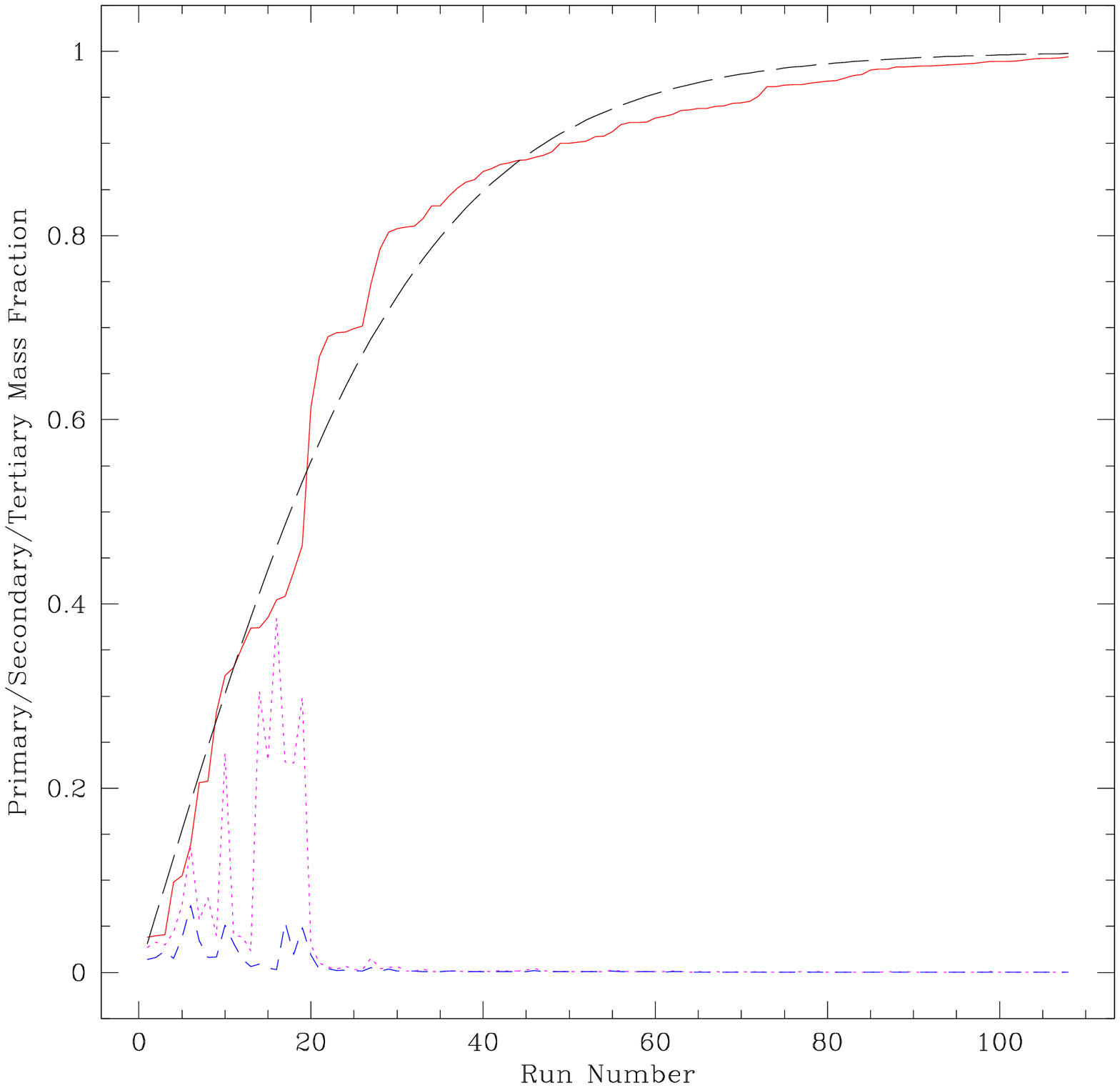}
\putfig{iso}{\textwidth}{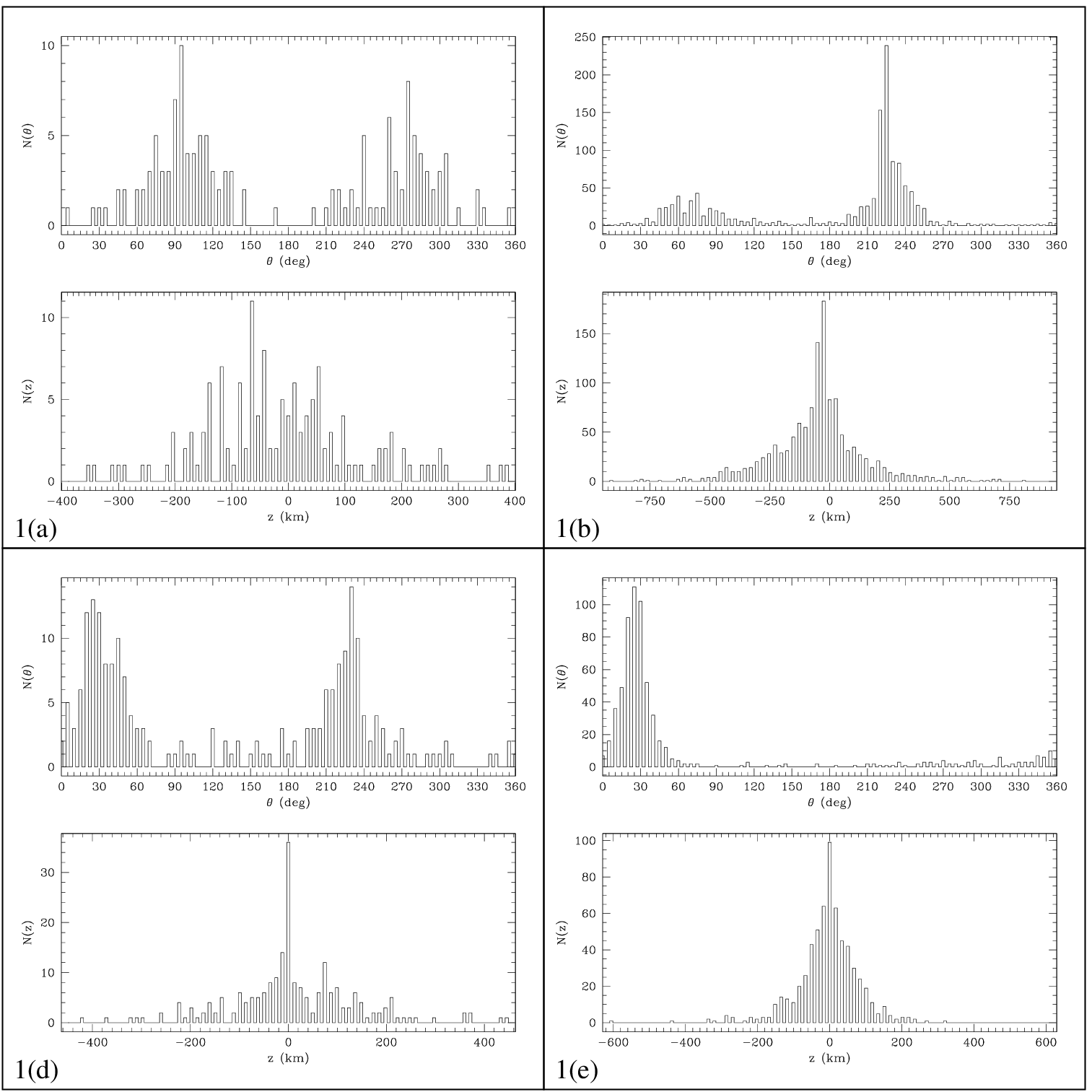}

\end{document}